\documentstyle[12pt]{article}
\begin{document}

\begin{titlepage}
\title{A Resonance Model for $\pi N \rightarrow Y K $ and
$\pi \Delta \rightarrow Y K$ Reactions
for Kaon Production in Heavy Ion Collisions}

\author{S. W. Huang\thanks{Supported by GSI under contract No. T\"U F\"as T},
K. Tsushima\thanks{Supported by DFG under contract No. Fa 67/14-1},
Amand Faessler, \\ E. Lehmann and Rajeev K. Puri\thanks{Supported
by BMFT under contract No. 06T\"U736} \\\\
Institut f\"ur Theoretische Physik, Universit\"at T\"ubingen\\
Auf der Morgenstelle 14, D-72076 T\"ubingen, F. R. Germany}
\date{\today}
\maketitle
\begin{abstract}

In a resonance model the reactions $\pi N \rightarrow Y K$ and
$\pi \Delta \rightarrow Y K$  are studied. For the reactions
$\pi N \rightarrow \Lambda K$
and $\pi \Delta \rightarrow \Lambda K$, the
resonances $N(1650) (J^P=\frac{1}{2}^-)$, $N(1710)
(\frac{1}{2}^+)$ and
$N(1720) (\frac{3}{2}^+)$ are included as intermediate
states. For the reactions $\pi N \rightarrow \Sigma K$ , the resonances
$N(1710) (\frac{1}{2}^+)$, $N(1720) (\frac{3}{2}^+)$ and
$\Delta (1920)(\frac{3}{2}^+)$ are considered, while
for the $\pi \Delta \rightarrow \Sigma K$ reactions the intermediate
resonances are $N(1710) (\frac{1}{2}^+)$ and $N(1720) (\frac{3}{2}^+)$.
Besides these resonances in the $s$-channel, the $t$-channel
$K^*(892)$-exchanges are also taken into account  as a smooth background.

The relevant coupling constants for the  meson-baryon
vertices are obtained (except for $\Delta (1920)$ ) from the
experimental decay branching ratios of the relevant resonances.
All isospin channels of the $\pi N \rightarrow Y K$ and
$\pi \Delta \rightarrow Y K$ cross sections are calculated.
By comparing the calculated results with the available experimental data,
we find that the total cross sections of the $\pi N \rightarrow Y K$
reactions can be explained by the resonance model. The
$\pi \Delta \rightarrow Y K$ cross sections, for which no
experimental data are available, are predicted theoretically.

Parametrizations of the calculated total cross sections for
all different isospin channels are
given for the use of kaon productions in heavy ion collisions.  The
differential cross sections are also studied.

\end{abstract}
\end{titlepage}

\section{Introduction}\

The purpose of this paper is to study
the $\pi N ({\rm and} \Delta) \rightarrow Y K$
($\Delta$ stands for $\Delta$(1232), $Y$ stands for either the $\Sigma$
or the $\Lambda$) reactions for kaon productions in heavy ion collisions.

 One of the main goals of intermediate energy heavy ion physics is
to determine the equation of state (EOS) of nuclear matter.
Theoretical studies show that the kaons produced in heavy ion
collisions are sensitive to the EOS \cite{aic}\cite{lan}.
To calculate the kaon production
in heavy ion  collisions one needs two ingredients \cite{cas1}\cite{li}:

1. A transport theory to describe the
evolution of the heavy ion colliding systems.

2. The elementary cross sections
of kaon production in the $BB$ and $\pi B$ collisions (here $B$
stands for either a nucleon $N$ or a $\Delta (1232)$).

Therefore, in order to obtain a definite conclusion about the EOS one needs
the reliable elementary cross sections for the kaon production
as well as the dynamical transport models.

One usually uses transport models such as BUU/VUU \cite{bert} and QMD
\cite{qmd}
to simulate the dynamical evolution of projectile and target systems. In
these models collision terms and the nuclear mean field are included. By
including the nuclear mean field one hopes to determine the EOS.
In the collision terms, the following reactions are included as
main processes:
$$B+B \rightarrow B + B \,\, {\rm (baryon \,elastic \,collision )} $$
$$N+N \rightarrow N + \Delta \,\, {\rm (\Delta \,production )}  $$
$$N+\Delta \rightarrow N+N \,\, {\rm  (\Delta \,absorption )}  $$
$$N+N \rightarrow N+N+ \pi \,\, {\rm  (pion \,production )}$$
$$\Delta \rightarrow N + \pi \,\, {\rm  (\Delta \,decay)} $$
$$\pi + N\rightarrow  \Delta \,\, {\rm  (pion \,absorption)} $$
\noindent
then, kaons can be produced mainly through
the following reactions:
$$B+B \rightarrow B+Y+K$$
and $$\pi +B \rightarrow Y+K.$$

The  Lorentz-invariant differential multiplicity
for kaons for a given impact parameter $b$ in heavy ion collisions is
connected with the elementary cross section  by
\begin{eqnarray}
E \frac{d^3N(b)}{d^3 p}& = & \sum _{BB_{coll}} \int \left (E'
\frac{d^3\sigma _{BB \rightarrow BYK} (\sqrt {s_{BB}}  )}{d^3 p'}
/\sigma _{BB}^{tot}(\sqrt { s_{BB} } ) \right )
[1-f({\bf r,p},t)] \frac{d\Omega }{4\pi}
\nonumber  \\ &   &
+ \sum _{\pi B_{coll}}  E''
\frac{d^3\sigma  _{\pi B\rightarrow Y K} (\sqrt {s_{\pi B}})}{d^3 p''}/
\sigma _{\pi B}^{tot}(\sqrt {s_{\pi B}}),
\label{basic1}
\end{eqnarray}

\noindent
where the primed and the double-primed quantities are in the
center-of-momentum (c.m.) frames of the
two colliding baryons ($BB$) and pion-baryon ($\pi B$), respectively.
The unprimed quantities are in the c.m. frame of the two nuclei.
$\sigma _{BB}^{tot}(\sqrt { s_{BB} } )$ and  $\sigma _{\pi B}^{tot}
(\sqrt {s_{\pi B} })$ are the total
cross sections for two baryons $(BB)$ and the pion-nucleon system
(${\pi N}$) as functions of c.m. energy $\sqrt {s_{BB} }$
and $\sqrt {s_{\pi B} }$,  respectively. The factor
$[1-f({\bf r,p},t)]$ stands for the Pauli blocking correction for the
final nucleon $N$ in the $BB \rightarrow NYK$ reaction due to
the surrounding nuclear medium.

The Lorentz-invariant differential kaon-production cross section in
heavy ion collisions is then given by:
\begin{equation}
{ E{d^3\sigma \over d^3 p} = 2\pi \int db ~b {E}{d^3N(b)
\over d^3  p } }.
\label{basic2}
\end{equation}
 From eq.(\ref{basic1}) we see that the elementary cross sections are
directly related to the kaon yields in heavy ion collisions.
Thus, it is important to have good elementary kaon-production cross
sections.

However, most elementary cross sections needed in studying heavy
ion collisions are not well known experimentally. Therefore one has to use
the available  experimental data as far as possible. In the case
when no experimental data are available one has to rely on
symmetry considerations and theoretical models to extrapolate the
available experimental data.

The elementary kaon-production cross section in $BB$ collisions
used in eq.(\ref{basic1})
are usually taken from Randrup and Ko\cite{ran}. In their works,
the $N\Delta \rightarrow
NYK$and  $\Delta \Delta \rightarrow NYK$ total cross sections are
obtained from  $N N \rightarrow NYK$ cross section by symmetry consideration
in isospin space, and the $\Delta$ is treated
as spin-$ \frac{1}{2} $ (not $\frac{3}{2}$) and isospin-$\frac{3}{2}$
particle. Moreover, the coupling constants relevant
to the $\Delta$ are assumed to be the same as for the nucleon $N$.

The elementary $K^+$-production cross section in $\pi B$ collisions
needed in eq.(\ref{basic1}) are usually taken from Cugnon and Lombard's
parametrization\cite{cug}.
In this parametrization
the isospin averaged cross sections are obtained from only three
available experimental data assuming proton and
neutron $(N=Z)$ symmetry.
This symmetry is, in principle, valid for the light nuclei,
and not appropriate for collisions between heavy nuclei
with a large neutron excess, i.e. $N> Z$. Furthermore the parametrization for
the reaction $\pi \Delta \rightarrow Y K$ is not given, hence the contribution
of this channel to kaon production in the heavy ion collision is unclear.

By using the elementary cross section of Randrup and Ko, one finds that
$N\Delta \rightarrow NYK$  gives the main contribution to
the total kaon yield in heavy ion collisions\cite{lan}.
However, the  elementary cross
section for this channel is experimentally unknown.
Thus a more sophisticated study of this
channel is still necessary. In order to determine the cross section
for the reaction $N\Delta \rightarrow NYK$,
one needs at first to study the reaction $\pi \Delta \rightarrow YK$
according to Randrup and Ko's model\cite{ran}\cite{fer}.

Since in heavy ion collisions many pions are produced, and the threshold of
$\pi N ({\rm or} \Delta) \rightarrow YK $ is lower than in
$BB \rightarrow NYK$,
the reaction $\pi N ({\rm or} \Delta) \rightarrow YK$
seems to be important as a secondary process
besides the  $BB \rightarrow NYK$ channel for kaon production.
According to a study of the Giessen group, the
$\pi N \rightarrow \Lambda K $ process is dominant for kaon
production in proton-nucleus reactions\cite{cas2}.

Recently, in the simulation codes of BUU and QMD the isospin
dependence has been considered. One distinguishes
the members of the nucleon isospin-doublet($p$ and $n$), of the pion
isospin-triplet($\pi^+$, $\pi^0$
and $\pi^-$) and of the delta
isospin-quartet ($\Delta^{++}$, $\Delta^+$, $\Delta^0$, and
$\Delta^-$)\cite{wol}\cite{hub}. This enables us to calculate the kaon
production
in an isospin dependent way,  hence one needs information on all the possible
isospin channels for kaon production. This means
 that the isospin-averaged elementary cross sections of $BB
\rightarrow BYK $ and $\pi N({\rm or} \Delta) \rightarrow YK $
can not be used for this purpose. For an isospin-dependent description
of kaon production, one needs all the isospin channels.
However, the cross sections for some isospin channels
are not available in  experiment.   Therefore one needs a theoretical
model to evaluate all the unknown isospin channels.

On the other hand, in order to study medium effects for
kaon production,
a dynamical model for the elementary cross sections is needed,
since one can not simply obtain the in-medium cross sections from a
parametrization of experimental data of the corresponding cross
sections in free space.

As a first step, we concentrate on the $\pi B \rightarrow Y K $
reactions. Once we know this cross sections
we can calculate the second term in eq.(\ref{basic1}). This channel is
also the basis to study $BB \rightarrow BYK$, since the
parameters of the $\pi B \rightarrow Y K $ reactions
 are ingredients for calculating the reaction
 $BB \rightarrow BYK$ in the present model.

 In this paper, we present results for all
 $\pi N ({\rm and} \Delta) \rightarrow Y K$
reactions using a resonance model. Part of this work has been
briefly reported
in \cite{tsu}. By a resonance model we mean that
the  $\pi N ({\rm and} \Delta) \rightarrow Y K $ reactions happen
via intermediate resonance states, which  decay both into
$\pi N ({\rm or} \Delta) $ and $Y K$.  The
coupling constants in the meson-baryon-resonances can be
determined from the relevant branching ratios of the resonances. The cross
sections  of $\pi N ({\rm and} \Delta) \rightarrow Y K $ are obtained
coherently from the square
of the sum of all resonance amplitudes. Besides these
$s$-channel resonances which give  peaks in the total
cross sections, the $t$-channel $K^*(892)$-exchange is also included.
It provides a smooth background.

\section{Experimental data for the resonance model} \

We study the  reactions  $\pi N ({\rm and} \Delta)  \rightarrow Y K$
for intermediate energies, i.e. for the range of the
$\pi N ({\rm and} \Delta)$  invariant mass
from the threshold for $YK$ production to about 3 GeV. According to
the  ``Review of Particle Properties''\cite{par}\cite{paro},
the relevant resonances to be included as intermediate states are
$N(1650) (J^P=\frac{1}{2}^-)$, $N(1710) (\frac{1}{2}^+)$,
$N(1720) (\frac{3}{2}^+)$ and $\Delta (1920)(\frac{3}{2}^+)$. \\

In this section we list the properties of the resonances
and of $K^*(892)$ relevant to our calculations.\\

(1) $ N(1650)\,\, L_{2I \, 2J}=S_{1 1}; I(J^P)=\frac{1}{2}(\frac{1}{2}^-) $:\\

Mass $m \approx 1650 $ MeV

Full width $\Gamma^{full} $=145 to 190 MeV $\approx$ 150 MeV

Partial decay modes and branching fractions:
\[  N(1650) \rightarrow \left\{
\begin{array}{ll}
                       N \pi & 60-80\%, {\rm we \, use} \, 70\%  \\
                       \Lambda K  &\approx 7\% \\
                       \Delta \pi  & < 10\% ,  {\rm we \, use}\, 5\%\\
                       \end{array}
                       \right.  \]

(2) $ N(1710) \,\, L_{2I\, 2J}=P_{1 1}; I(J^P)=\frac{1}{2}(\frac{1}{2}^+) $:\\

Mass $ m \approx 1710 $ MeV

Full width $\Gamma^{full} $=50 to 250 MeV $\approx$ 100 MeV

Partial decay modes and branching fractions:
\[  N(1710) \rightarrow \left\{
\begin{array}{ll}
                       N \pi & 10-20\%,  {\rm we \, use} \, 15\%       \\
                       \Lambda K &5-25\%,  {\rm we \, use}  \,  15\%  \\
                       \Delta \pi  & 10-25\% , {\rm we \, use} \, 17.5\% \\
                       \Sigma K  & 2-10\% ,  {\rm we \, use} \, 6\% \\
                       \end{array}
                       \right.  \]

(3) $ N(1720) \,\, L_{2I \, 2J}=P_{1 3}; I(J^P)=\frac{1}{2}(\frac{3}{2}^+) $:\\

Mass $ m \approx 1720 $ MeV

Full width $\Gamma^{full} $=100 to 200 MeV $\approx$ 150 MeV

Partial decay modes and branching fractions:
\[  N(1720) \rightarrow \left\{
\begin{array}{ll}
                       N \pi & 10-20\%,  {\rm we \, use} \, 15\%       \\
                       \Lambda K &3-10\%,  {\rm we \, use}  \,  6.5\%  \\
                       \Delta \pi  & 5-15\% ,  {\rm we \, use} \, 10\% \\
                       \Sigma K  & 2-5\% ,  {\rm we \, use} \, 3.5 \% \\
                       \end{array}
                       \right.  \]

(4) $\Delta(1920) \,\, L_{2I\, 2J}=P_{3 3}; I(J^P)=\frac{3}{2}(\frac{3}{2}^+)
$:\\

Mass $ m \approx 1920  $ MeV

Full width $\Gamma^{full} $=150 to 300 MeV $\approx$ 200 MeV

Partial decay modes and branching fractions:
\[  \Delta(1920) \rightarrow \left\{
\begin{array}{ll}
                       N \pi & 5-20\%,  {\rm we \, use} \, 12.5\%       \\
                       \Sigma K  & 1-3\% ,  {\rm we \, use} \, 2 \% \\
                       \end{array}
                       \right.  \]

(5) $ K^*(892) \,\,  I(J^P)=\frac{1}{2}({1}^-) $ :\\

Mass $ m \approx 892 $ MeV

Full width $\Gamma ^{full} $=49.8 MeV

Partial decay modes and branching fractions:
$$  K^*(892) \rightarrow   K \pi ~ 100 \% $$

\section{Formalism}\

In the present study, the mesons, baryons and baryon-resonances are
treated as fundamental fields. Due to the fact that the
relevant hadrons have finite sizes, form factors are included at
every  vertex.

In section 3.1 to 3.4 we give the  effective interaction Lagrangians,
and the formulas for the transition amplitudes and cross sections.
Throughout this work SU(2) symmetry in isospin space is assumed.
In detail we treat
 $\pi N \rightarrow \Lambda K $,
 $\pi N \rightarrow \Sigma K $,
 $\pi \Delta \rightarrow \Lambda K $,
 $\pi \Delta \rightarrow \Sigma K $ in the following subsections $\S$3.1,
 $\S$3.2, $\S$3.3 and $\S$3.4 separately.

\subsection{$\pi+N \rightarrow \Lambda +K $ reactions}\

 From the experimental data listed in  $\S$ 2, three
resonances $N(1650) (\frac{1}{2}^-)$, $N(1710) (\frac{1}{2}^+)$ and
$N(1720) (\frac{3}{2}^+)$ contribute to the
$\pi N \rightarrow \Lambda K$ reactions. In addition $K^*(892)$-exchange
can also be  responsible for the same reaction.
The relevant Feynman diagrams are shown in fig. 1:

The effective interaction Lagrangians needed in
describing the vertices are given by:

\begin{equation}
{\cal L}_{\pi N N(1650)} =
-g_{\pi N N(1650)}
\left( \bar{N}(1650)  \vec\tau N \cdot \vec\phi
+ \bar{N} \vec\tau  N(1650) \cdot \vec\phi\,\, \right),
\label{pinn1650}
\end{equation}

\begin{equation}
{\cal L}_{\pi N N(1710)} =
-ig_{\pi N N(1710)}
\left( \bar{N}(1710) \gamma_5 \vec\tau N \cdot \vec\phi
+ \bar{N} \vec\tau \gamma_5 N(1710) \cdot \vec\phi\,\, \right),
\label{pinn1710}
\end{equation}

\begin{equation}
{\cal L}_{\pi N N(1720)} =
\frac{g_{\pi N N(1720)}}{m_\pi}
\left( \bar{N}^\mu(1720) \vec\tau N \cdot \partial_\mu \vec\phi
+ \bar{N} \vec\tau N^\mu(1720) \cdot \partial_\mu \vec\phi \,
\right),
\label{pinn1720}
\end{equation}

\begin{equation}
{\cal L}_{K \Lambda  N(1650)} =
-g_{K \Lambda  N(1650)}
\left( \bar{N}(1650)   \Lambda K
+\bar{K}\bar{\Lambda}  N(1650)  \,\, \right),
\label{klan1650}
\end{equation}

\begin{equation}
{\cal L}_{K \Lambda  N(1710)} =
-ig_{K \Lambda  N(1710)}
\left( \bar{N}(1710) \gamma_5  \Lambda K
+\bar{K}\bar{\Lambda} \gamma_5  N(1710)  \,\, \right),
\label{klan1710}
\end{equation}

\begin{equation}
{\cal L}_{K \Lambda  N(1720)} =
\frac{g_{K \Lambda N(1720)}}{m_K}
\left( \bar{N}^{\mu}(1720)   \Lambda \partial _{\mu}  K
+  (\partial_{\mu} \bar{K}) \bar{\Lambda} N^{\mu}(1720)\,\, \right),
\label{klan1720}
\end{equation}

\begin{equation}
{\cal L}_{K^*(892)  \Lambda  N} = - g_{K^*(892)  \Lambda N }
\left( \bar{N} \gamma^\mu  \Lambda K^*_\mu(892)
+ \frac{\kappa} {m_N + m_\Lambda} \bar{N} \sigma^{\mu \nu}
\Lambda \partial_\mu K^*_\nu(892)
+ {\rm h. c.} \right),
\label{kslan}
\end{equation}

\begin{equation}
{\cal L}_{K^*(892) K \pi} = i f_{K^*(892) K \pi}
\left( \bar{K} \vec\tau K^*_\mu(892) \cdot
\partial^\mu \vec\phi
- (\partial^\mu \bar{K}) \vec\tau K^*_\mu(892) \cdot \vec\phi \,\right)
+ {\rm h. c.},
\label{kskpi}
\end{equation} \\

In the above Lagrangians, $N$, $N(1650)$, $N(1710)$, $\Lambda$ stand for
spin-$\frac{1}{2}$ Dirac spinors to describe the corresponding particles.
Spin-$\frac{3}{2}$ Rarita-Schwinger fields
$\psi^\mu = N^\mu(1720)$ with mass $m$
satisfy the equations\cite{tak}:
\begin{eqnarray}
 & ( i \gamma \cdot \partial - m ) \psi^\mu=0, \label{R-S1}   \\
 & \gamma_\mu \psi^\mu=0,                      \label{R-S2}   \\
 & \partial_\mu \psi^\mu=0,                   \label{R-S3}
\end{eqnarray}
and $\vec \phi $ stands for the pion field. In isospin space
nucleon fields are
expressed by $N = \left( p\, , \, n \right)^T$, where the superscript
``$T$'' means the transposition operation, similar expressions
can also be written for the nucleon resonances.
The meson field operators appearing in eqs. (\ref{pinn1650})-(\ref{kskpi})
are related to the physical representations as follows,
$K=\left(K^+\,, K^0 \right)^T,\,\,$
$\bar{K}=\left(K^-\, ,\bar{K^0} \right),\,$
$K^*_\mu(892)=\left(K^*_\mu(892)^+,\, K^*_\mu(892)^0 \right)^T,\,$
$\bar{K}^*_\mu(892)=\left(K^*_\mu(892)^-,\,
\bar{K}^*_\mu(892)^0 \right),\,$
$\pi^{\pm}=\frac{1}{\sqrt{2}}(\phi_1 \mp i \phi_2),\,\, \pi^0=\phi_3.\,\,$

For the propagators $iS_F(p)$ of the spin-$\frac{1}{2}$ and
the propagators $iG^{\mu \nu}(p)$ of the spin-$\frac{3}{2}$
resonances\cite{ben} we use

\begin{equation}
iS_F(p) =i \frac{\gamma \cdot p + m}{p^2 - m^2 +
im\Gamma^{full}}\, \,\,\,  {\rm and}
\end{equation}

\begin{equation}
iG^{\mu \nu}(p) =i \frac{P^{\mu \nu}(p)}{p^2 - m^2 +
im\Gamma^{full}}\,,
\end{equation}

with
\begin{equation}
P^{\mu \nu}(p) = - (\gamma \cdot p + m)
\left[ g^{\mu \nu} - \frac{1}{3} \gamma^\mu \gamma^\nu
- \frac{1}{3 m}( \gamma^\mu p^\nu - \gamma^\nu p^\mu)
- \frac{2}{3 m^2} p^\mu p^\nu \right],   \label{pmunu}
\end{equation}
\newline
respectively, where $m$ and $\Gamma^{full}$ stand for the mass
and the full decay width of the corresponding baryon resonance.

According to the Lagrangians given by eqs. (\ref{pinn1650})-(\ref{kskpi}),
the amplitudes for the $\pi N \rightarrow \Lambda K$ reactions are given by
\begin{equation}
{\cal M}_{\pi^0 p \rightarrow \Lambda K^+}=
-{\cal M}_{\pi^0 n \rightarrow \Lambda K^0}
= {\cal M}_{a1}
+ {\cal M}_{b1}
+ {\cal M}_{c1}
+ {\cal M}_{d1}
\label{nlamamp1}
\end{equation}
\begin{equation}
{\cal M}_{\pi^+ n \rightarrow \Lambda K^+}=
{\cal M}_{\pi^- p \rightarrow \Lambda K^0}
=\sqrt {2} \left( {\cal M}_{a1}
+ {\cal M}_{b1}
+ {\cal M}_{c1}
+ {\cal M}_{d1} \right)
\label{nlamamp2}
\end{equation}

\noindent
where the  resonance amplitudes ${\cal M}_{a1}$
($N(1650)$ intermediate state), ${\cal M}_{b1}$
($N(1710)$ intermediate state), ${\cal M}_{c1}$ ($N(1720)$
intermediate state) and the $K^*(892)$-exchange
amplitude ${\cal M}_{d1}$ are given by
\begin{equation}
{\cal M}_{a1}
=\frac{ g_{ \pi N N(1650) } g_{K \Lambda N(1650) }\bar{u}_\Lambda
(p_\Lambda ,s_\Lambda)\,( \gamma \cdot p + m_{N(1650)})
u_N(p_N,s_N) }{p^2 - m_{N(1650)}^2 + i m_{N(1650)}
\Gamma_{N(1650)}^{full}} \enspace ,
\label{a1}
\end{equation}
\begin{equation}
{\cal M}_{b1}
=\frac{ -g_{ \pi N N(1710) } g_{K \Lambda N(1710) }
\bar{u}_\Lambda (p_\Lambda ,s_\Lambda) \gamma_5 \,(
\gamma \cdot p + m_{N(1710)})
\gamma_5 u_N(p_N,s_N)}{p^2 - m_{N(1710)}^2 + i m_{N(1710)}
\Gamma_{N(1710)}^{full}} \enspace ,
\label{b1}
\end{equation}
\begin{equation}
{\cal M}_{c1}
=\frac{-g_{\pi N N(1720)}g_{K \Lambda N(1720)}{{p_K}_\mu {p_\pi}_\nu
\bar{u}_\Lambda (p_\Lambda  ,s_\Lambda)\, P_{N(1720)}^{\mu \nu}(p)\,
u_N(p_N, s_N), }  }{m_\pi m_K
{(p^2 - m_{N(1720)}^2 + i m_{N(1720)}
\Gamma_{N(1720)}^{full}} ) }\,
\enspace
\label{c1}
\end{equation}
$${\cal M}_{d1} =
\frac{f_{K^*(892) K \pi} g_{K^*(892) \Lambda N}}
{(p_\Lambda - p_N)^2 - m_{K^*(892)}^2} \enspace
\bar{u}_\Lambda (p_\Lambda, s_\Lambda)\,
\left[ \gamma_\mu - i \frac{\kappa}{ m_N + m_\Lambda }
\sigma_{\alpha \mu} (p_\Lambda - p_N)^\alpha \right] \hspace{2cm}
$$
\begin{equation}
\hspace{5cm} \cdot (p_\pi + p_K)_\nu \left( g^{\mu \nu} -
\frac{(p_\Lambda - p_N)^\mu (p_\Lambda - p_N)^\nu}{m_{K^*(892)}^2} \right)\,
u_N(p_N,s_N).
\label{d1}
\end{equation}

\noindent
Here $u_N(p_N,s_N)$ and $u_\Lambda(p_\Lambda,s_\Lambda)$ in eqs.
(\ref{a1})-(\ref{d1}) are the spinors
of the nucleon $N$ and the lambda $\Lambda$, with four-momentum $p_N$,
spin $s_N$ and four-momentum $p_\Lambda$, $s_\Lambda$,  respectively.
$p$ is the four-momentum of the resonance. $p_\pi$, $p_K$ are the
four-momenta of the pion and the kaon, respectively.

We obtain the absolute values of the effective coupling constants
by equating the experimental branching ratios given in $\S$ 2 with the
theoretical calculations given in appendix  $\S$ 5.1 .

Form factors (denoted by $F$ and
$F_{K^*(892) K \pi}$ below)
are introduced. They reflect the finite size  of the hadrons.
These form factors must be multiplied to each vertex of the interactions
\footnote{For simplicity form factors are not explicitly given in
the amplitudes. They have to be included at each vertex.}.
Thus, the coupling constants are obtained from the branching
ratios in the rest frame of the resonances.

For the form factors in baryon-baryon-meson vertices we use

\begin{equation}
F(\mid {\vec q} \mid)=\frac{\Lambda_C^2}{\Lambda_C^2 + {\vec q}\,^2},
\label{form1}
\end{equation}
with $\mid {\vec q}\mid$ being the magnitude of three-momentum
${\vec q}$ of the mesons, and  $\Lambda_C$ being the cut off parameter.

For the $K^*(892)$-$K$-$\pi$ vertex  we adopt the form
factor of ref. \cite{gob}.

\begin{equation}
F_{K^*(892) K \pi}(\mid \frac{1}{2}(\vec{p_K}-\vec{p_\pi}) \mid)
= C \mid \frac{1}{2}(\vec{p_K}-\vec{p_\pi}) \mid
\exp\left( - \beta \mid\frac{1}{2}(\vec{p_K}-\vec{p_\pi})\mid^2 \right).
\label{form2}
\end{equation}

In evaluating the cross sections in
the c.m. frame of the $\pi N$ system, each coupling constant
$g_{PBB^*}$, $g_{K^*(892)\Lambda N}$ and $f_{K^*(892)K\pi}$
appearing in  eqs. (\ref{a1})-(\ref{d1}) must be replaced by
$g_{PBB^*} \rightarrow g_{PBB^*}F(\mid{\vec q}_P\mid)$,
$g_{K^*(892)\Lambda N} \rightarrow g_{K^*(892)\Lambda N}
F(\mid\vec{q}_\Lambda-\vec{q}_N \mid)$
and
$f_{K^*(892)K\pi} \rightarrow f_{K^*(892)K\pi}
F_{K^*(892)K\pi}( |\frac{1}{2}(\vec{p_K}-\vec{p_\pi})| )$. Here
$B^*$ stands for the nucleon resonances.
$B$ indicates a baryon ($N(938)$ or $\Lambda$) and $P$ stands
for pseudoscalar-meson ($\pi$ or  $K$), respectively.

The (spin unpolarized) differential cross section for
$\pi^0 p \rightarrow \Lambda K^+ $ in the c.m. frame of $\pi^0$ $p$ is
given by
\begin{equation}
\frac{d\sigma ( \pi^0 p \rightarrow \Lambda K^+) }{d\Omega}=
\frac{1}{(8 \pi \sqrt s)^2 }\frac{|{\vec q_f}|}{|{\vec q_i}|}
\frac{1}{2} \sum_{all \, spins} |{\cal M}_{\pi^0 p \rightarrow \Lambda K^+}|^2,
\label{cross1}
\end{equation}
where ${\vec q_f}$ stands in the c.m. frame for the outgoing three-momenta,
and
${\vec q_i}$  for the incoming  three-momenta of the two particles.
We have, in the c.m. frame of $\pi N$ system,

\begin{equation}
|{\vec q_f}|=|{\vec p_\Lambda}|=|{\vec p_K}|=\frac{\lambda^{\frac{1}{2}}
(s,m_\Lambda^2,m_K^2)}{2\sqrt{s}}
\end{equation}
and
\begin{equation}
|{\vec q_i}|=|{\vec p_N}|=|{\vec p_\pi}|=\frac{\lambda^{\frac{1}{2}}
(s,m_N^2,m_{\pi}^2)}{2\sqrt{s}}.
\end{equation}
Here $s=(p_N+p_\pi)^2=p^2$ is the Mandelstam variable, and $\lambda (x,y,z)$
is defined by\cite{itz}
\begin{equation}
\lambda (x,y,z)\equiv x^2+y^2+z^2-2xy-2xz-2yz.
\label{tri}
\end{equation}

The total cross section for the reaction
$\pi^0 p \rightarrow \Lambda K^+ $ reads

\begin{equation}
\sigma (\pi^0 p \rightarrow \Lambda K^+)=\int d \Omega \frac{ d\sigma
(\pi^0 p \rightarrow \Lambda K^+) }{d\Omega}.
\label{total1}
\end{equation}

Similar expressions for other isospin channels for $K^+$ and
$K^0$-production can be obtained in the same way
as in eqs. (\ref{nlamamp1}) and (\ref{nlamamp2}). \\



%

\subsection{$\pi+N \rightarrow \Sigma +K $ reactions}\

Besides $K^*$-exchange, there are three
resonances contributing to the $\pi N \rightarrow \Sigma K$
reactions (see $\S$2). They are $N(1710) (\frac{1}{2}^+)$,
$N(1720) (\frac{3}{2}^+)$ and $\Delta (1920)(\frac{3}{2}^+)$.
The Feynman diagrams are shown in fig. 2.

In addition to eqs.(\ref{pinn1710})-(\ref{pinn1720}) and eq.(\ref{kskpi}),
the interaction Lagrangians needed in
describing the vertices relevant for the reactions
$\pi N \rightarrow \Sigma K$ are given by:

\begin{equation}
{\cal L}_{\pi N \Delta(1920)} =
\frac{g_{\pi N \Delta(1920)}}{m_\pi}
\left( \bar{\Delta}^\mu (1920) \overrightarrow{\cal I} N \cdot
\partial_\mu \vec\phi + \bar{N} {\overrightarrow{\cal I}}^\dagger
\Delta^\mu(1920) \cdot \partial_\mu \vec\phi \, \right),
\label{pind1920}
\end{equation}
\begin{equation}
{\cal L}_{K \Sigma N(1710)} =
-ig_{K \Sigma N(1710)}
\left( \bar{N}(1710) \gamma_5 \vec\tau \cdot \overrightarrow\Sigma K
+ \bar{K} \overrightarrow{\bar \Sigma} \cdot \vec\tau
\gamma_5 N(1710) \right),
\label{ksin1710}
\end{equation}
\begin{equation}
{\cal L}_{K \Sigma N(1720)} =
\frac{g_{K \Sigma N(1720)}}{m_K}
\left( \bar{N}^\mu (1720) \vec\tau \cdot \overrightarrow\Sigma
\partial_\mu K + (\partial_\mu \bar{K}) \overrightarrow{\bar \Sigma}
\cdot \vec\tau N^\mu (1720) \right),
\label{ksin1720}
\end{equation}
\begin{equation}
{\cal L}_{K \Sigma \Delta(1920)} =
\frac{g_{K \Sigma \Delta(1920)}}{m_K}
\left( \bar{\Delta}^\mu(1920) \overrightarrow{\cal I}
\cdot \overrightarrow\Sigma \partial_\mu K
+ (\partial_\mu \bar{K}) \overrightarrow{\bar \Sigma} \cdot
{\overrightarrow{\cal I}}^\dagger \Delta^\mu (1920) \right),
\label{ksid1920}
\end{equation}
\begin{equation}
{\cal L}_{K^*(892) \Sigma N}=-g_{K^*(892) \Sigma N}
\left( \bar{N} \gamma^\mu \vec\tau \cdot \overrightarrow\Sigma K^*_\mu(892)
+ \frac{\kappa}{m_N + m_\Sigma} \bar{N} \sigma^{\mu \nu}
\vec\tau \cdot \overrightarrow\Sigma \partial_\mu K^*_\nu(892)
+ {\rm h. c.} \right),
\label{kssin}
\end{equation}
\vspace{1mm}
\noindent

In the eqs. above $\Sigma$ stands for the spin-$\frac{1}{2}$ Dirac spinor
of the $\Sigma$ particle. $\Delta^\mu (1920)$ stands for the
spin-$\frac{3}{2}$ Rarita-Schwinger field of $\Delta (1920)$,
which satisfies the equations (\ref{R-S1})-(\ref{R-S3}).
$\kappa$ is the ratio of the tensor coupling
constant to the vector coupling constant.
$\vec{\cal I}$ is the isospin transition operator defined by
\begin{eqnarray*}
\overrightarrow{\cal I}_{Mm} &= \displaystyle{\sum_{\ell=\pm1,0}}
(1 \ell \frac{1}{2} m | \frac{3}{2} M)
\hat{e^*}_{\ell},
\end{eqnarray*}
\noindent
with
\[ \left\{
\begin{array}{ll}
\hat{e}_1=-{\frac{1}{\sqrt2}}(1,i,0)   \\
\hat{e}_0=                   (0,0,1)   \\
\hat{e}_{-1}={\frac{1}{\sqrt2}}(1,-i,0)  \\
\end{array}
\right.  \]

\noindent
and $\vec \tau$ are the Pauli matrices.
$\Delta ^\mu (1920)=$$\left( \Delta^\mu (1920)^{++},\,\,
\Delta^\mu(1920)^+,\,\,
\Delta^\mu(1920)^0,\,\, \Delta^\mu(1920)^- \right)^T$.
The $\Sigma$ field operators appearing in eqs. (\ref{ksin1710})-(\ref{kssin})
are related to the physical representations by
$\Sigma^{\pm}=\frac{1}{\sqrt{2}} (\Sigma_1 \mp i \Sigma_2),\,\,
\Sigma^0=\Sigma_3\,\,.$

The required amplitudes for the cross section are given by
\begin{equation}
{\cal M}_{\pi^+ p \rightarrow \Sigma^+ K^+}=
{\cal M}_{\pi^- n \rightarrow \Sigma^- K^0}=
\left(
{\cal M}_{c2}+
2{\cal M}_{d2}
\right) ,
\label{nsiamp1}
\end{equation}
\begin{equation}
{\cal M}_{\pi^- p \rightarrow \Sigma^- K^+}=
{\cal M}_{\pi^+ n \rightarrow \Sigma^+ K^0}=
2\left(
{\cal M}_{a2}+
{\cal M}_{b2}+\frac{1}{6}
{\cal M}_{c2}
\right) ,
\label{nsiamp2}
\end{equation}
\begin{equation}
{\cal M}_{\pi^+ n \rightarrow \Sigma^0 K^+}=
{\cal M}_{\pi^0 p \rightarrow \Sigma^+ K^0}=
\sqrt 2\left(
{\cal M}_{a2}+
{\cal M}_{b2}-\frac{1}{3}
{\cal M}_{c2}-{\cal M}_{d2}
\right) ,
\label{nsiamp3}
\end{equation}
\begin{equation}
{\cal M}_{\pi^0 n \rightarrow \Sigma^- K^+}=
{\cal M}_{\pi^- p \rightarrow \Sigma^0 K^0}=
-\sqrt 2\left(
{\cal M}_{a2}+
{\cal M}_{b2}-\frac{1}{3}
{\cal M}_{c2}-{\cal M}_{d2}
\right) ,
\label{nsiamp4}
\end{equation}
\begin{equation}
{\cal M}_{\pi^0 p \rightarrow \Sigma^0 K^+ }=
{\cal M}_{\pi^0 n \rightarrow \Sigma^0 K^0 }=
{\cal M}_{a2}+
{\cal M}_{b2}+\frac{2}{3}
{\cal M}_{c2}+{\cal M}_{d2},
\label{nsiamp5}
\end{equation}
\noindent
where the single resonance amplitudes ${\cal M}_{a2}$ (with
the $N(1710)$ resonance),  ${\cal M}_{b2}$ ($N(1720)$),
${\cal M}_{c2}$ ($\Delta(1920)$ ), and the $K^*(892)$-exchange
amplitude ${\cal M}_{d2}$ in the above equations are given by
{\small
\begin{equation}
{\cal M}_{a2}=
\frac{- g_{\pi N N(1710)}g_{K \Sigma N(1710)}\bar{u}_\Sigma
(p_\Sigma, s_\Sigma)\, \gamma_5\, (\gamma \cdot p + m_{N(1710)})\,
\gamma_5\, u_N(p_N,s_N)}
{p^2 - m_{N(1710)}^2 + i m_{N(1710)}
\Gamma_{N(1710)}^{full}} \enspace  ,
\label{a2}
\end{equation}
}
{\small
\begin{equation}
{\cal M}_{b2}=
\frac{g_{\pi N N(1720)}g_{K \Sigma N(1720)}{p_K}_\mu
{p_\pi}_\nu\bar{u}_\Sigma (p_\Sigma, s_\Sigma)\,
P_{N(1720)}^{\mu \nu}(p)\, u_N(p_N, s_N)}{m_\pi m_K
(p^2 - m_{N(1720)}^2 + i m_{N(1720)}\Gamma_{N(1720)}^{full})}\,,
\label{b2}
\end{equation}
}
{\small
\begin{equation}
{\cal M}_{c2}=
\frac{g_{\pi N \Delta(1920)}g_{K \Sigma \Delta(1920)}
{p_K}_\mu {p_\pi}_\nu \bar{u}_\Sigma (p_\Sigma,s_\Sigma )\,
P_{\Delta(1920)}^{\mu \nu}(p)\, u_N(p_N,s_N)}{m_\pi m_K
(p^2 - m_{\Delta(1920)}^2 + i m_{\Delta(1920)}
\Gamma_{\Delta(1920)}^{full})}\, ,
\label{c2}
\end{equation}
}
$$
{\cal M}_{d2}=
\frac{f_{K^*(892) K \pi} g_{K^*(892) \Sigma N}}
{(p_\Sigma - p_N)^2 - m_{K^*(892)}^2} \enspace
\bar{u}_\Sigma (p_\Sigma,s_\Sigma)\,
\left[ \gamma_\mu - i \frac{\kappa}{ m_N + m_\Sigma }
\sigma_{\alpha \mu} (p_\Sigma - p_N)^\alpha \right] \hspace{3cm}
$$
\begin{equation}
\hspace{5cm} \cdot (p_\pi + p_K)_\nu \left( g^{\mu \nu} -
\frac{(p_\Sigma - p_N)^\mu (p_\Sigma - p_N)^\nu}{m_{K^*(892)}^2} \right)\,
u_N(p_N,s_N).
\label{d2}
\end{equation}

Form factors are inserted at each vertex in eqs. (\ref{a2})-(\ref{d2})
in the same way as in $\S$ 3.1.

The differential cross section for $\pi^+ p \rightarrow \Sigma^+ K^+$
in the c.m. frame of $\pi^+ p$ is given by

\begin{equation}
\frac{d\sigma (\pi^+ p \rightarrow \Sigma^+ K^+)}{d\Omega}=
\frac{1}{(8 \pi \sqrt s)^2}\frac{|{\vec q_f}|}{|{\vec q_i}|}
\frac{1}{2} \sum_{all \, spins} |{\cal M}_{\pi^+ p \rightarrow
\Sigma^+ K^+}|^2,
\label{cross2}
\end{equation}
\noindent
where ${\vec q_f}$ stands for the outgoing three-momenta of
the $\Sigma^+$ and the $K^+$ in the c.m. system, and
${\vec q_i}$ stands for the  incoming  three-momenta of the
$\pi^+$ and the $p$.
We have, in the c.m. frame of $\pi N$ system,

\begin{equation}
|{\vec q_f}|=|{\vec p_\Sigma}|=|{\vec p_K}|=\frac{\lambda^{\frac{1}{2}}
(s,m_\Sigma^2,m_K^2)}{2\sqrt{s}}
\end{equation}
and
\begin{equation}
|{\vec q_i}|=|{\vec p_N}|=|{\vec p_\pi}|=\frac{\lambda^{\frac{1}{2}}
(s,m_N^2,m_{\pi}^2)}{2\sqrt{s}}.
\end{equation}

The total cross section for $\pi^+ p \rightarrow \Sigma^+ K^+$ reads

\begin{equation}
\sigma (\pi^+ p \rightarrow \Sigma^+ K^+)=\int d \Omega \frac{ d\sigma
(\pi^+ p \rightarrow \Sigma^+ K^+)}{d\Omega}.
\label{total2}
\end{equation}

The same expressions as eqs. (\ref{cross2}) and (\ref{total2})
for other isospin dependent reactions can be obtained by
eqs. (\ref{nsiamp2})-(\ref{nsiamp5}).
%
%
%
 From eqs. (\ref{nsiamp3}) and eq.(\ref{nsiamp4}), we derive

\begin{equation}
\frac{d\sigma (\pi^0 n\rightarrow \Sigma^- K^+)}{{d\Omega}}=
\frac{d\sigma (\pi^+ n\rightarrow \Sigma^0 K^+)}{d\Omega}.
\end{equation}

The $K^0$ and the $K^+$ production have the same amplitudes
given by eqs. (\ref{nsiamp1})-(\ref{nsiamp5}),
for example, from eq. (\ref{nsiamp1}), one finds:

\begin{equation}
\frac{d\sigma (\pi^- n\rightarrow \Sigma^- K^0)}{{d\Omega}}=
\frac{d\sigma (\pi^+ p\rightarrow \Sigma^+ K^+)}{d\Omega}.
\end{equation} \\





\subsection{$\pi+\Delta(1232) \rightarrow \Lambda +K $ reactions}\

 From the experimental data listed in  $\S$ 2, three
resonances  $N(1650) (\frac{1}{2}^-)$, $N(1710) (\frac{1}{2}^+)$,
$N(1720) (\frac{3}{2}^+)$ contribute to the
$\pi \Delta(1232)  \rightarrow \Lambda K $ reactions.
Due to isospin conservation  $K^*(892)$-exchange can not
contribute to the  $\pi \Delta(1232) \rightarrow \Lambda K $ reactions.
The relevant Feynman diagrams are shown in fig. 3:

Besides eqs.(\ref{klan1650})-(\ref{klan1720}), the following three
interaction Lagrangian are needed in describing the  vertices of the
$\pi \Delta \rightarrow \Lambda K$ reactions:
\begin{equation}
{\cal L}_{\pi \Delta N(1650)} =
i \frac{g_{\pi \Delta N(1650)}}{m_\pi}
\left( \bar{N}(1650) \gamma_5 {\overrightarrow{\cal I}}^\dagger
\Delta^\mu \cdot \partial_\mu \vec\phi
+ \bar{\Delta}^\mu \overrightarrow{\cal I} \gamma_5 N(1650) \cdot
\partial_\mu \vec\phi\, \right),
\label{pidn1650}
\end{equation}
\begin{equation}
{\cal L}_{\pi \Delta N(1710)} =
\frac{g_{\pi \Delta N(1710)}}{m_\pi}
\left( \bar{N}(1710) {\overrightarrow{\cal I}}^\dagger
\Delta^\mu \cdot \partial_\mu \vec\phi
+ \bar{\Delta}^\mu \overrightarrow{\cal I} N(1710) \cdot
\partial_\mu \vec\phi\, \right),
\label{pidn1710}
\end{equation}
\begin{equation}
{\cal L}_{\pi \Delta N(1720)} =
- i g_{\pi \Delta N(1720)}
\left( \bar{N}^\mu(1720) \gamma_5
{\overrightarrow{\cal I}}^\dagger \Delta_\mu  \cdot \vec\phi
+ \bar{\Delta}_\mu \overrightarrow{\cal I} \gamma_5 N^\mu(1720)
\cdot \vec\phi \, \right).
\label{pidn1720}
\end{equation}
The amplitudes for the reactions are given by

\begin{equation}
{\cal M}_{\pi^- \Delta^{++} \rightarrow  \Lambda K^+}=
-{\cal M}_{\pi^+ \Delta^{-} \rightarrow  \Lambda K^0}=
-\left({\cal M}_{a3}+
{\cal M}_{b3}+
{\cal M}_{c3}\right),
\label{dlamamp1}
\end{equation}
\begin{equation}
{\cal M}_{\pi^0 \Delta^+ \rightarrow  \Lambda K^+}=
{\cal M}_{\pi^0 \Delta^0 \rightarrow  \Lambda K^0}=
\sqrt{\frac{2}{3}} \left({\cal M}_{a3}+{\cal M}_{b3}+
{\cal M}_{c3}\right),
\label{dlamamp2}
\end{equation}
\begin{equation}
{\cal M}_{\pi^+ \Delta^0 \rightarrow  \Lambda K^+}=
-{\cal M}_{\pi^- \Delta^+ \rightarrow  \Lambda K^0}=
\frac{1}{\sqrt{3}} \left({\cal M}_{a3}+
{\cal M}_{b3}+
{\cal M}_{c3}\right).
\label{dlamamp3}
\end{equation}\\
\noindent
Here the amplitudes ${\cal M}_{a3}$ (with $N(1650)$ as the
intermediate state), ${\cal M}_{b3}$($N(1710)$) and ${\cal M}_{c3}$($N(1720)$)
corresponding to each of the diagrams (a), (b) and (c) given in fig. 3 are
given by(the coefficients in eqs. (\ref{dlamamp1})-(\ref{dlamamp3}),
are from isospin):

\begin{equation}
{\cal M}_{a3}= \frac{- g_{\pi \Delta N(1650)}g_{K \Lambda N(1650)}}{m_\pi}
\,\,\frac{\,{p_\pi}_{\mu}\, \bar{u}_\Lambda (p_\Lambda,s_\Lambda)\,
(\gamma \cdot p + m_{N(1710)})\,\gamma_5\,
u_\Delta^\mu(p_\Delta,s_\Delta)}
{p^2 - m_{N(1650)}^2 + i m_{N(1650)}
\Gamma_{N(1650)}^{full}} ,
\label{a3}
\end{equation}
\begin{equation}
{\cal M}_{b3}=
\frac{- g_{\pi \Delta N(1710)}g_{K \Lambda N(1710)}}{m_\pi}
\,\,\frac{\,{p_\pi}_{\mu}\, \bar{u}_\Lambda (p_\Lambda,s_\Lambda)\, \gamma_5\,
(\gamma \cdot p + m_{N(1710)})\, u_\Delta^\mu(p_\Delta,s_\Delta)}
{p^2 - m_{N(1710)}^2 + i m_{N(1710)}
\Gamma_{N(1710)}^{full}} ,
\label{b3}
\end{equation}
\begin{equation}
{\cal M}_{c3} = \frac{g_{\pi \Delta N(1720) }g_{K \Lambda N(1720)}}{m_K}
\,\,\frac{\,{p_K}_{\mu}\, \bar{u}_\Lambda (p_\Lambda,s_\Lambda)\,
P_{N(1720)}^{\mu \nu}(p)\, \gamma_5\,
u_\Delta {}_\nu (p_\Delta,s_\Delta)}
{p^2 - m_{N(1720)}^2 + i m_{N(1720)}
\Gamma_{N(1720)}^{full}} .
\label{c3}
\end{equation}
\noindent
The  Rarita-Schwinger spinor-vector
$u_\Delta^\mu(p_\Delta,s_\Delta)$ of the delta's
with momentum $p_\Delta$ and spin $s_\Delta$,
$u_\Delta^\mu(p_\Delta,s_\Delta)$  satisfies in momentum space
the eqs. (\ref{R-S1})-(\ref{R-S3}).

\begin{eqnarray}
&(\gamma \cdot p_\Delta -m_\Delta)u_\Delta^\mu(p_\Delta,s_\Delta)=0,
\label{R-S4} \\
&\gamma_\mu u_\Delta^\mu(p_\Delta, s_\Delta)=0,       \label{R-S5} \\
&p_\Delta {}_\mu u_\Delta^\mu(p_\Delta, s_\Delta)=0.  \label{R-S6}
\end{eqnarray}

Form factors are inserted at each vertex in eqs. (\ref{a3})-(\ref{c3})
in the same way as in $\S$ 3.1 and $\S$ 3.2.

The differential cross section for $\pi^- \Delta^{++} \rightarrow
\Lambda K^+ $ in the c.m. frame of $\pi^- \Delta^{++}$ is given by

\begin{equation}
\frac{d\sigma (\pi^- \Delta^{++} \rightarrow  \Lambda K^+ ) }{d\Omega}=
\frac{1}{(8 \pi \sqrt s)^2 }\frac{|{\vec q_f}|}{|{\vec q_i}|}
\frac{1}{4} \sum_{all \, spins}
|{\cal M}_{\pi^- \Delta^{++} \rightarrow  \Lambda K^+}|^2,
\label{cross3}
\end{equation}

where
\begin{equation}
|{\vec q_f}|=|{\vec p_\Lambda }|=|{\vec p_K}|=\frac{\lambda^{\frac{1}{2}}
(s,m_\Lambda^2,m_K^2)}{2\sqrt{s}}
\end{equation}
and
\begin{equation}
|{\vec q_i}|=|{\vec p_\Delta }|=|{\vec p_\pi}|=\frac{\lambda^{\frac{1}{2}}
(s,m_{\Delta}^2,m_{\pi}^2)}{2\sqrt{s}}
\end{equation}
with
\begin{equation}
s=(p_\Delta+p_\pi)^2=p^2.
\end{equation}

The total cross section for $\pi^- \Delta^{++} \rightarrow
\Lambda K^+ $ reads

\begin{equation}
\sigma (\pi^- \Delta^{++} \rightarrow  \Lambda K^+ )=
\int d \Omega \frac{ d\sigma
(\pi^- \Delta^{++} \rightarrow  \Lambda K^+ )}{d\Omega}.
\label{total3}
\end{equation}

The relationships between other isospin channels  and the above channel
can be easily obtained from eqs.(\ref{dlamamp1})-(\ref{dlamamp3}).\\

%



\subsection{$\pi+\Delta(1232) \rightarrow \Sigma +K $ reactions}\

 According to the data listed in $\S$ 2, two resonances contribute
to the  $\pi \Delta \rightarrow \Sigma K $ reactions:
 $N(1710) (\frac{1}{2}^+)$ and
$N(1720) (\frac{3}{2}^+)$. Further the t-channel $K^*(892)$-exchange
is also relevant for these reactions. The Feynman diagrams are
shown in fig. 4.

In addition to Lagrangians given by eqs. (\ref{ksin1710})-(\ref{ksin1720}),
(\ref{pidn1710})-(\ref{pidn1720}) and eq. (\ref{kskpi}),
one more interaction Lagrangian needed in
describing  the $\pi \Delta \rightarrow \Sigma K$ reactions is given by
\begin{equation}
{\cal L}_{K^*(892) \Sigma \Delta} = - i g_{K^*(892) \Sigma \Delta}
\left( \bar{K}^*_\mu(892) {\overrightarrow{\cal I}}^\dagger
\gamma_5 \Delta^\mu
+ \bar{\Delta}^\mu \gamma_5 \overrightarrow{\cal I}
K^*_\mu(892) \, \right).
\end{equation}
The amplitudes for the $\pi \Delta \rightarrow \Sigma K$ reactions
are given by:

\begin{equation}
{\cal M}_{\pi^- \Delta^{++} \rightarrow \Sigma^0 K^+}=
{\cal M}_{\pi^+ \Delta^{-} \rightarrow \Sigma^0 K^0}=
-({\cal M}_{a4}+{\cal M}_{b4}) ,
\label{dsiamp1}
\end{equation}
\begin{equation}
{\cal M}_{\pi^- \Delta^+ \rightarrow  \Sigma^- K^+} =
-{\cal M}_{\pi^+ \Delta^0 \rightarrow  \Sigma^+ K^0} =
-\sqrt{\frac{2}{3}} ({\cal M}_{a4}+{\cal M}_{b4})  ,
\label{dsiamp2}
\end{equation}
\begin{equation}
{\cal M}_{\pi^0 \Delta^{++} \rightarrow  \Sigma^+ K^+} =
{\cal M}_{\pi^0 \Delta^{-} \rightarrow  \Sigma^- K^0} =
-{\cal M}_{c4} ,
\label{dsiamp3}
\end{equation}
\begin{equation}
{\cal M}_{\pi^0 \Delta^+ \rightarrow \Sigma^0 K^+}=
-{\cal M}_{\pi^0 \Delta^0 \rightarrow \Sigma^0 K^0}=
\sqrt{\frac{2}{3}} ({\cal M}_{a4}+{\cal M}_{b4}+{\cal M}_{c4}) ,
\label{dsiamp4}
\end{equation}
\begin{equation}
{\cal M}_{\pi^0 \Delta^0 \rightarrow \Sigma^- K^+}=
{\cal M}_{\pi^0 \Delta^+ \rightarrow \Sigma^+ K^0}=
\frac{1}{\sqrt{3}} (2{\cal M}_{a4}+2{\cal M}_{b4}+{\cal M}_{c4}),
\label{dsiamp5}
\end{equation}
\begin{equation}
{\cal M}_{\pi^+ \Delta^+ \rightarrow  \Sigma^+ K^+} =
-{\cal M}_{\pi^- \Delta^0 \rightarrow  \Sigma^- K^0} =
-\sqrt{\frac{2}{3}} {\cal M}_{c4} ,
\label{dsiamp6}
\end{equation}
\begin{equation}
{\cal M}_{\pi^+ \Delta^0 \rightarrow \Sigma^0 K^+}=
{\cal M}_{\pi^- \Delta^+ \rightarrow \Sigma^0 K^0}=
\frac{1}{\sqrt{3}}
({\cal M}_{a4}+{\cal M}_{b4}+2{\cal M}_{c4}  )  ,
\label{dsiamp7}
\end{equation}
\begin{equation}
{\cal M}_{\pi^+ \Delta^- \rightarrow \Sigma^- K^+}=
-{\cal M}_{\pi^- \Delta^{++} \rightarrow \Sigma^+ K^0}=
\sqrt{2}
({\cal M}_{a4}+{\cal M}_{b4}+{\cal M}_{c4})  .
\label{dsiamp8}
\end{equation}

\noindent
The amplitudes ${\cal M}_{a4}$ (with  $N(1710)$ as an
intermediate resonance),
${\cal M}_{b4}$($N(1720)$) and ${\cal M}_{c4}$
($K^*(892)$-exchange) corresponding to
diagrams (a), (b) and (c) in fig. 4 are given as follows(isospin factors
are included by the coefficients in  eqs. (\ref{dsiamp1})-(\ref{dsiamp8})):
\begin{equation}
{\cal M}_{a4} = \frac{-g_{\pi \Delta N(1710)}g_{K \Sigma N(1710)}}{m_\pi}
\,\,\frac{\,{p_\pi}_{\mu}\, \bar{u}_\Sigma (p_\Sigma,s_\Sigma)\, \gamma_5\,
(\gamma \cdot p + m_{N(1710)})\, u_\Delta^\mu(p_\Delta,s_\Delta)}
{p^2 - m_{N(1710)}^2 + i m_{N(1710)}
\Gamma_{N(1710)}^{full}} ,
\label{a4}
\end{equation}
\begin{equation}
{\cal M}_{b4} = \frac{g_{\pi \Delta N(1720)}g_{K \Sigma N(1720)}}{m_K}
\,\,\frac{\,{p_K}_{\mu}\, \bar{u}_\Sigma (p_\Sigma,s_\Sigma)\, P_{N(1720)}^{\mu
\nu}(p)\, \gamma_5\,
u_\Delta {}_\nu (p_\Delta,s_\Delta)}
{p^2 - m_{N(1720)}^2 + i m_{N(1720)}
\Gamma_{N(1720)}^{full}} ,
\label{b4}
\end{equation}
\begin{equation}
{\cal M}_{c4} = \frac{i f_{K^*(892) K \pi}g_{K^*(892) \Sigma \Delta}}
{(p_\Sigma - p_\Delta)^2 - m^2_{K^*(892)}}
\,\, \bar{u}_\Sigma (p_\Sigma,s_\Sigma) \gamma_5
u_\Delta^\mu(p_\Delta,s_\Delta)
(p_\pi + p_K)^\nu
\left( g_{\mu \nu}
- \frac{(p_\Sigma - p_\Delta)_\mu (p_\Sigma - p_\Delta)_\nu}
{m^2_{K^*(892)}} \right).
\label{c4}
\end{equation}
Note that the $N(1650)$ resonance
gives contributions to the $\pi \Delta \rightarrow \Lambda K$
reactions but does not give contributions to the
$\pi \Delta \rightarrow \Sigma K$ reactions.

Form factors are inserted at each vertex in eqs. (\ref{a4})-(\ref{c4})
in the same way as in $\S$ 3.1, $\S$ 3.2 and  $\S$ 3.3 .

The differential cross section for the reaction
$\pi^- \Delta^{++} \rightarrow \Sigma^0 K^+ $ in the c.m. frame
of $\pi \Delta$ reads

\begin{equation}
\frac{d\sigma (\pi^- \Delta^{++} \rightarrow \Sigma^0 K^+  ) }{d\Omega}=
\frac{1}{(8 \pi \sqrt s)^2 }\frac{|{\vec q_f}|}{|{\vec q_i}|}
\frac{1}{4} \sum_{all \, spins}
|{\cal M}_{\pi^- \Delta^{++} \rightarrow \Sigma^0 K^+}|^2,
\end{equation}
%
%
%
%
%

where
\begin{equation}
|{\vec q_f}|=|{\vec p_\Sigma }|=|{\vec p_K}|=\frac{\lambda^{\frac{1}{2}}
(s,m_\Sigma^2,m_K^2)}{2\sqrt{s}}
\end{equation}
and
\begin{equation}
|{\vec q_i}|=|{\vec p_\Delta }|=|{\vec p_\pi}|=\frac{\lambda^{\frac{1}{2}}
(s,m_{\Delta}^2,m_{\pi}^2)}{2\sqrt{s}}.
\end{equation}

Other isospin channels and $K^0$ productions can be obtained by
eqs.(\ref{dsiamp1})-(\ref{dsiamp8}). \\

\section{Results and Discussions}\

In this section we give the results and discuss the cross
sections of the reactions given in section 3. We treat in detail
the reactions
 $\pi N \rightarrow \Lambda K $,
 $\pi N \rightarrow \Sigma K $,
 $\pi \Delta \rightarrow \Lambda K $,
 $\pi \Delta \rightarrow \Sigma  K $ in the  subsections $\S$4.1,
 $\S$4.2, $\S$4.3 and $\S$4.4, respectively. In each subsection
we first discuss the total and then the differential cross sections.

\subsection{$\pi+N \rightarrow \Lambda +K $ reactions}\

The total cross section for $\pi^0+p\rightarrow \Lambda +K^+$
is given by eq. (\ref{total1}).
Since all the resonances contributing to the
$\pi+N \rightarrow \Lambda +K $ reactions
have isospin $\frac{1}{2}$, different isospin channels of the
$\pi+N \rightarrow \Lambda +K $ reactions differ only by an
overall factor.
The cross section for
$\pi^- +p \rightarrow \Lambda + K^0 $ is two times the one of
$\pi^0+p \rightarrow \Lambda +K^+$  according to eqs. (\ref{nlamamp1}) and
(\ref{nlamamp2}).
The amplitudes appearing in eqs. (\ref{nlamamp1}) and (\ref{nlamamp2})
are given by eqs. (\ref{a1})-(\ref{d1}).

In evaluation of these amplitudes we need eight coupling
constants (i.e. $g_{\pi N N(1650)}$, $g_{\pi N N(1710)}$, $g_{\pi N N(1720)}$,
 $g_{K \Lambda N(1650)}$, $g_{K \Lambda N(1710)}$, $g_{K \Lambda N(1720)}$,
 $g_{K^*(892) \Lambda N }$, $f_{K^*(892) K \pi }$ )
and  cutoff  parameters. From the eight coupling constants,
the absolute values of seven can
be determined by equating eqs. (\ref{ratio511})-(\ref{ratio517})
given in appendix 5.1 with the experimental branching ratios
listed in $\S$2. One coupling constant, i.e. $g_{K^*(892)\Lambda N}$,
is treated as a free parameter in the present work. As mentioned in $\S$ 3,
we include form factors in the decay branching
ratios as well as in the scattering amplitudes throughout this
work (see also eqs. (\ref{form1}) and (\ref{form2})).
By choosing the cutoff values as :
$\Lambda_c=0.8$ GeV for $N(1650)$, $N(1710)$ and $N(1720)$,
we obtain the coupling constants in table 1.
For the $K^*(892)\Lambda N$ vertex we use $g_{K^*(892)\Lambda N}=0.45$,
$\Lambda_c=1.2$ GeV. The cut off parameters $C$ and $\beta$
in $F_{K^*(892) K \pi}$ are  $C = 2.72$ fm and
$\beta = 8.88 \times 10^{-3}$ fm$^2$ used in ref. \cite{gob},
and the resulting coupling constant $f_{K^*(892) K \pi}$ is
also shown in table 1.

According to  eqs.(\ref{nlamamp1}), (\ref{nlamamp2}) and (\ref{cross1})
we sum the amplitudes of the different resonances, and
obtain the cross sections by squaring the sum of
the amplitudes. Thus we include all interference terms.(The
problem with the signs of the coupling constants is discussed below.)

First, we show the total cross section for the reaction
$\pi^- +p \rightarrow \Lambda + K^0 $ without interference terms
in fig. 5(a).  The experimental data are from ref\cite{bal}.
Since the experimental branching ratios
can determine the square of the coupling constants only,
the relative signs of the interference terms are not determined
(in the present case  $2^{4-1}=8$  possibilities).
We have tested all these possibilities. Fig. 5(b) shows
the result with the interference terms with one choice of the relative signs.
In selecting this sign combination we choose the one which gives
the best results compared with the experimental data for both the
total and the differential cross section,
we then choose this sign combination in all further investigations.
We also found that other relative sign combinations
give a better total cross section, however do not give a
good differential cross section.

A single resonance or $K^*(892)$-exchange can not
explain the experimental data.
The contribution from $N(1650)$ is larger than from the other resonances.
But one must include all the resonances from the particle booklet which
decay  into $\pi N$ and $K \Lambda$
for the description of kaon production.

The $s$-channel resonances gives a
Breit-Wigner form as expected, whereas the $t$-channel $K^*$-exchange is
necessary to explain the long tail in the total cross section.

Figs. 6(a), 6(b) and 6(c) show the results for differential
cross sections of the reaction $\pi^- +p \rightarrow \Lambda + K^0 $
at the pion beam momentum 0.980 GeV/c, 1.13 GeV/c and 1.455 GeV/c
respectively.
The experimental data of Knasel {\it et al.} \cite{kna75} and Saxon
{\it et al.} \cite{sax80} are shown
in the figures with error bars.  One sees that the theoretical results
are in roughly
good agreement with the experimental data.

We mention here
that in the work of Brown {\it et al.}\cite{bro}, the isospin-averaged
total cross section of $\pi +N \rightarrow \Lambda + K $ is
parametrized in the Breit-Wigner form by including the $N(1650)$,
the $N(1710)$, the $N(1720)$. The interference terms between the resonances
are not taken into account. Since the amplitudes
are not given, they can not study the differential cross sections.

Our  parametrization for the $\pi^- p \rightarrow \Lambda K^0$
cross section is given by
\begin{equation}
\sigma(\pi^- p \rightarrow \Lambda K^0)
 = \frac{0.007665 (\sqrt{s}-1.613)^{0.1341}}{(\sqrt{s}-1.720)^2+0.007826}
\hspace{0.5cm} {\rm mb},
\label{param1}
\end{equation}
for $\sqrt{s}$  smaller than the $\Lambda  K$ production
threshold of 1.613 GeV, the above cross section is zero.
Other isospin channels can be related to the reaction of
$\pi^- p \rightarrow \Lambda K^0$  by eqs. (\ref{nlamamp1}) and
(\ref{nlamamp2}).

\subsection{$\pi+N \rightarrow \Sigma +K$  reactions}\

The total and differential cross sections of the $\pi^+ +p
\rightarrow \Sigma^+ + K^+$ reaction are given by
eqs. (\ref{total2}) and  (\ref{cross2}), respectively.
The expressions for other isospin dependent reactions can be
written in the same way.  The amplitudes for each channel are
shown in eqs. (\ref{a2})-(\ref{d2}).

In evaluation of the amplitudes we need eight coupling
constants (i.e. $g_{\pi N N(1710)}$, $g_{\pi N N(1720)}$,
$g_{\pi N \Delta(1920)}$,
 $g_{K \Sigma N(1710)}$, $g_{K \Sigma N(1720)}$, $g_{K \Sigma \Delta (1920)}$,
 $g_{K^*(892) \Sigma N}$ and $f_{K^*(892) K \pi }$)
and  cutoffs. Among these eight coupling constants, three
(i.e. $g_{\pi N N(1710)}$,  $g_{\pi N N(1720)}$ and  $f_{K^*(892) K \pi}$)
are already given in table 1.
The coupling constant and the cut off parameter of the
$\Sigma-N-K^*(892)$ vertex is assumed to be the same as
$\Lambda-N-K^*(892)$ vertex in order to introduce a minimum of free parameters,
i.e. $g_{K^*(892) \Sigma N}$=$g_{K^*(892) \Lambda N}$=0.45.
The other four coupling constants (i.e. $g_{\pi N \Delta(1920)}$,
$g_{K \Sigma N(1710)}$, $g_{K \Sigma N(1720)}$ and
$g_{K \Sigma \Delta (1920)}$)
are determined by comparing eqs.
(\ref{ratio521})-(\ref{ratio524}) given in appendix 5.2 with the
experimental branching ratios  listed in $\S$ 2. (The values of  $g_{\pi N
\Delta(1920)}$ and
$g_{K \Sigma \Delta (1920)}$ obtained in this way will be called
``set 2'', the reason will get clear below.)
By using the cutoff value  $\Lambda_C=0.8$ GeV  for the $N(1710)$,
the $N(1720)$ as in $\S$ 4.1, and $\Lambda_C=0.5$ GeV for the
$\Delta(1920)$, these four coupling constants are shown in table 2.

Unlike  the $\pi+N \rightarrow \Lambda +K $ reactions described
in $\S$ 3.1 and $\S$ 4.1, the resonances
$N(1710)$, $N(1720)$, $\Delta (1920)$ contribute to these  reactions,
due to the mixture of the isospin-$\frac{1}{2}$ of the nucleon resonances
(N(1710) and N(1720)),
and isospin-$\frac{3}{2}$ of the $\Delta(1920)$ resonance,
the different isospin channels of the $\pi+N \rightarrow \Sigma +K$
reactions differ not only by a overall factor but also in shape.
This point can be seen from  eqs. (\ref{nsiamp1})-(\ref{nsiamp5}).
In these equations the amplitudes ${\cal M}_{a2}$, ${\cal M}_{b2}$,
${\cal M}_{c2}$ and ${\cal M}_{d2}$ contribute to different
isospin channels of the  $\pi+N \rightarrow \Sigma +K$ reactions
with different weights.  Namely the $\Delta (1920)$ and
$K^*(892)$-exchange distinguish different isospin channels of
$\pi+N \rightarrow \Sigma +K $ reactions. The role of the $\Delta (1920)$
in the  $\pi+N \rightarrow \Sigma +K $ reactions was studied
in our previous work \cite{tsu}.

For the  $\pi+N \rightarrow \Sigma +K $ reactions
three experimental data sets exist in the literature:

(1). $\pi^+ +p \rightarrow \Sigma^+ + K^+$ ,

(2). $\pi^- +p \rightarrow \Sigma^- + K^+$ ,

(3). $\pi^- +p \rightarrow \Sigma^0 + K^0$.

All these three channels have indeed different shapes, i.e.
a different energy dependence of the total cross section.
First, we discuss the reaction  $\pi^+ +p \rightarrow \Sigma^+ +
K^+ $. In this channel the initial  and the final states have
charge two. Therefore
only the doubly charged  resonance $\Delta(1920)^{++}$ and
the $K^*(892)$-exchange can contribute
(see eq. (\ref{nsiamp1})).
The result is shown in
the dashed line of fig. 7(a).  The cross section is considerably
underestimated.
This means that the contributions from other
$\Delta$ resonances with  masses near 1.9 GeV
can not be neglected\footnote{In the compilation
of "Review of Particle Properties"
(Phys. Rev. D50 1173 (1994)) there are $\Delta(1900), \Delta(1905),
\Delta(1910), \Delta(1930), \Delta(1940), \Delta(1950)$
contribute to this reaction, however the value of branching ratios
are still not yet determined.}.
Thus we consider the $\Delta(1920)$ as an
effective  resonance that simulates also other $\Delta$ resonances
in this energy region that contribute to this reaction.

In this work we scale the two coupling constants
$g_{K \Sigma \Delta (1920)}$ and $g_{\pi N \Delta(1920)}$
of the $\Delta(1920)$
to account for other $\Delta$ resonance contributions
by fitting the
$ \pi^+ +p \rightarrow \Sigma^+ +K^+ $ experimental
data. Their values are also shown in table 2.  These two scaled
coupling constants $g_{K \Sigma \Delta
(1920)}$ and $g_{\pi N \Delta(1920)}$ are also used in all other
channels of our
calculations. All other coupling constants remain  unchanged.
The calculations with these two scaled coupling
constants  are called ``set 1".  The results with these two
coupling constants unscaled as determined from the branching ratios
are also shown in the fig. 7  for reference ( ``set 2'').
We believe that the ``set 1"  parameters produce reliable results and
should be used for kaon production in heavy ion collisions.

Fig. 7(b) shows the result of the $\pi^- +p \rightarrow \Sigma^- + K^+$
reaction.  We find that ``set 1"  results are in good agreement with the data,
especially for the second peak, while ``set 2'' can not reproduce
the second peak. This means that the two scaled coupling constants
give a reliable description also of these data, although fitted
to an other reaction.

According to eq. (\ref{nsiamp3}) and eq. (\ref{nsiamp4}) the cross sections
for
$\pi^0 +n \rightarrow \Sigma^- + K^+$ and
$\pi^+ +n \rightarrow \Sigma^0 + K^+$ reactions are the same as for
the $\pi^- +p \rightarrow \Sigma^0 + K^0$ reaction.
We thus compare the calculations of the reactions
$\pi^0 +n \rightarrow \Sigma^- + K^+$ and
 $\pi^+ +n \rightarrow \Sigma^0 + K^+$  with the
experimental data for
 $\pi^- +p \rightarrow \Sigma^0 + K^0$  in fig. 7(c).
 From  fig. 7(c) we see that these reactions are also in good
agreement with the experimental data.
 From figs. 7(a)-7(c) we see that the experimental data for
different isospin
channels have  indeed a different shape.  The theoretical
calculations  reproduce these different shapes.
For the reaction $\pi^0 +p \rightarrow \Sigma^0 + K^+$ no data are
available. Since this cross
section is important for heavy ion collisions, we give
the results for ``set 1'' parameters in eq(\ref{prmtpips4}).

Since in the present approach we scaled the two coupling
constants to take into account other $\Delta$ resonances around
1.9 GeV, we can not demand that the differential cross section can also
be reproduced. The differential cross section  strongly
depends on the quantum numbers of the resonance. The differential
cross section can be studied in the future if the branching ratios
of the relevant resonances  are well determined.

In order to use these cross sections in heavy ion collisions, we
parameterize all the channels based on ``set 1'' parameters as follows:

They are:
\begin{equation}
\sigma(\pi^+ p \rightarrow \Sigma^+ K^+)
 = \frac{0.03591 (\sqrt{s}-1.688)^{0.9541}}{(\sqrt{s}-1.890)^2+0.01548}
 + \frac{0.1594 (\sqrt{s}-1.688)^{0.01056}}{(\sqrt{s}-3.000)^2+0.9412}
\hspace{0.5cm} {\rm mb},
\end{equation}
\begin{equation}
\sigma(\pi^- p \rightarrow \Sigma^- K^+)
 = \frac{0.009803 (\sqrt{s}-1.688)^{0.6021}}{(\sqrt{s}-1.742)^2+0.006583}
 + \frac{0.006521 (\sqrt{s}-1.688)^{1.4728}}{(\sqrt{s}-1.940)^2+0.006248}
\hspace{0.5cm} {\rm mb},
\end{equation}
\begin{equation}
\sigma(\pi^+ n \rightarrow \Sigma^0 K^+)
= \sigma(\pi^0 n \rightarrow \Sigma^- K^+)
= \frac{0.05014 (\sqrt{s}-1.688)^{1.2878}}{(\sqrt{s}-1.730)^2+0.006455}
\hspace{0.5cm} {\rm mb}, \hspace{3em}
\end{equation}
\begin{equation}
\sigma(\pi^0 p \rightarrow \Sigma^0 K^+)
 = \frac{0.003978 (\sqrt{s}-1.688)^{0.5848}}{(\sqrt{s}-1.740)^2+0.006670}
 + \frac{0.04709 (\sqrt{s}-1.688)^{2.1650}}{(\sqrt{s}-1.905)^2+0.006358}
\hspace{0.5cm} {\rm mb}, \label{prmtpips4}
\end{equation}
\\
where, all parametrizations given above should be understood
to be zero below the threshold $\sqrt{s} \leq 1.688$ GeV.
These parametrizations are especially useful for kaon production simulation
codes for those channels where no experimental data are available.

\subsection{$\pi+\Delta (1232) \rightarrow \Lambda +K$  reactions}\

The total and differential cross sections of the reaction
$\pi^- \Delta^{++} \rightarrow  \Lambda K^+$  are given by
eqs. (\ref{total3}) and (\ref{cross3}), respectively.
The same expressions for other isospin dependent reactions can be
written in the same way.  The amplitudes for each channels are given
by eqs. (\ref{a3})-(\ref{c3}).

We need six coupling constants (i.e.  $g_{\pi \Delta N(1650)}$,
$g_{\pi \Delta N(1710)}$, $g_{\pi \Delta N(1720)}$,
$g_{K \Lambda N(1650)}$, $g_{K \Lambda N(1710)}$ and
$g_{K \Lambda N(1720)}$ ), three of them (i.e. $g_{K \Lambda N(1650)}$,
$g_{K \Lambda N(1710)}$ and $g_{K \Lambda N(1720)}$) are given
in $\S$ 4.1 and listed in table 1. Three other coupling
constants are obtained from
the experimental data listed in $\S$ 2 and eqs. (\ref{ratio531})-
(\ref{ratio533})  given in appendix 5.3.  By choosing the cutoff values as :
$\Lambda_c=0.8$ GeV for $N(1650)$, $N(1710)$ and $N(1720)$,
i.e. in the same way as in $\S$ 4.1 and  $\S$ 4.2, we obtain these three
coupling constants as listed in table 3.

Since all the resonances that contribute to the reactions
$\pi+\Delta \rightarrow \Lambda +K$
have isospin-$\frac{1}{2}$, the different isospin
dependent reaction channels have all the same shape (see eqs.
(\ref{dlamamp1})-(\ref{dlamamp3}). This is the same as for the
$\pi+N \rightarrow \Lambda +K$ reactions.

We find that the contribution from the $N(1710)$ resonance is larger (about
$60\%$ of the total cross section at peak position) than from other
resonances.

For this reaction one has four possible relative sign combinations.
Eq.(\ref{prmtpidla}) shows the total cross section of the reaction $\pi^-
\Delta^{++}
\rightarrow  \Lambda K^+$ without interference terms. A
careful study did show that the  interference terms do not
affect the final results appreciably for this reaction.
\\
\begin{equation}
\sigma(\pi^- \Delta^{++} \rightarrow \Lambda K^+)
 = \frac{0.006545 (\sqrt{s}-1.613)^{0.7866}}{(\sqrt{s}-1.720)^2+0.004852}
\hspace{0.5cm} {\rm mb}, \label{prmtpidla}
\end{equation}
This parametrization  should be understood to be zero below
the threshold ($\sqrt{s} \leq 1.613$ GeV).
This parametrization is useful for codes which simulate
kaon production since no experimental data are available.

We find that the angular distribution  $\pi^- \Delta^{++} \rightarrow
\Lambda K^+ $ is almost constant for all  bombarding energies.

\subsection{$\pi+\Delta(1232) \rightarrow \Sigma +K$  reactions}\

We need six coupling constants (i.e.  $g_{\pi \Delta N(1710)}$,
$g_{\pi \Delta N(1720)}$, $g_{K \Sigma N(1710)}$,
$g_{K \Sigma N(1720)}$,  $g_{K^*(892) \Sigma \Delta}$,
$f_{K^*(892) K \pi}$), three of them ( $g_{K \Sigma N(1710)}$,
$g_{K \Sigma N(1720)}$, $f_{K^*(892) K \pi}$) are given in $\S$ 4.2
and table 2. Two coupling constant ($g_{\pi \Delta N(1710)}$,
$g_{\pi \Delta N(1720)}$) are given in  $\S$ 4.3 and
table 3.  One remaining coupling constant  $g_{K^*(892) \Sigma \Delta}$
is assumed to be  $\sqrt3 g_{K^*(892) \Sigma N}$ where $\sqrt3$
comes from the different normalization of the operator $\vec \tau$
for isospin-$\frac{1}{2}$ and $\vec {\cal I} $ for isospin-$\frac{3}{2}$.
The cut off parameter $\Lambda_c$ for vertex $K^*(892) \Sigma \Delta$
is the same as $K^*(892) Y N$ used in $\S$ 4.1 and $\S$ 4.2 .

According to the different reaction amplitudes given by
eqs. (\ref{dsiamp1})-(\ref{dsiamp8}), one sees that the
${K^*(892)}$-exchange distinguishes  channels with
different isospin projections. Hence, among eight possible reaction
channels there are five different independent
reactions which differ not only by a overall factors, they are

(1).  $\pi^- \Delta^{++} \rightarrow \Sigma^0 +K^+$,

(2).  $\pi^0 \Delta^{0}  \rightarrow \Sigma^- +K^+$,

(3).  $\pi^+ \Delta^{0}  \rightarrow \Sigma^0 +K^+$ ,

(4).  $\pi^+ \Delta^{-}  \rightarrow \Sigma^- +K^+$ and

(5).  $\pi^0 \Delta^{++} \rightarrow \Sigma^+ +K^+$.

We find that the cross section of the reaction
$\pi^0 \Delta^{++} \rightarrow \Sigma^+
K^+$, to which only the ${K^*(892)}$-exchange
contribute (see eq.(\ref{dsiamp3})),
is negligible (smaller than 0.01 mb at its maximum value).
We plot the
reactions (1)-(3) in fig. 8. The reaction
 $\pi^+ \Delta^{-}  \rightarrow \Sigma^- K^+$ has a maximum
cross section of 0.15 mb with the peak position located at $\sqrt
s=1.75 $ GeV.
Due to the small contributions from the ${K^*(892)}$-exchange, the shape from
the different reactions of (1) -(4) do not differ much especially
at the peak positions.

We also find that the differential cross sections of the reactions
$\pi+\Delta(1232) \rightarrow \Sigma +K$  are almost
constant as a function of $cos\theta_{cm}$ for all  beam energies.

Now, we are in a position to give parametrizations of
total cross sections
$\pi^- \Delta^{++} \rightarrow \Sigma^0 K^+$,
$\pi^0 \Delta^0 \rightarrow \Sigma^- K^+$,
$\pi^+ \Delta^0 \rightarrow \Sigma^0 K^+$,
and $\pi^+ \Delta^- \rightarrow \Sigma^- K^+$
which are sufficient to describe all channels given in
eqs. (\ref{dsiamp1})-(\ref{dsiamp8}). Channel (5) is omitted,
since only $K^*(892)$-exchange contributes, which is
negligible as mentioned above.
Since the signs of interference terms cannot be fixed by
experimental data, we parameterize the results
obtained without interference terms.
\\
\begin{equation}
\sigma(\pi^- \Delta^{++} \rightarrow \Sigma^0 K^+)
= \frac{0.004959 (\sqrt{s}-1.688)^{0.7785}}{(\sqrt{s}-1.725)^2+0.008147}
\hspace{0.5cm} {\rm mb}, \label{prmtpids1}
\end{equation}
\begin{equation}
\sigma(\pi^0 \Delta^0 \rightarrow \Sigma^- K^+)
= \frac{0.006964 (\sqrt{s}-1.688)^{0.8140}}{(\sqrt{s}-1.725)^2+0.007713}
\hspace{0.5cm} {\rm mb}, \label{prmtpids2}
\end{equation}
\begin{equation}
\sigma(\pi^+ \Delta^0 \rightarrow \Sigma^0 K^+)
= \frac{0.002053 (\sqrt{s}-1.688)^{0.9853}}{(\sqrt{s}-1.725)^2+0.005414}
+ \frac{0.3179 (\sqrt{s}-1.688)^{0.9025}}{(\sqrt{s}-2.675)^2+44.88}
\hspace{0.5cm} {\rm mb},  \label{prmtpids3}
\end{equation}
\begin{equation}
\sigma(\pi^+ \Delta^- \rightarrow \Sigma^- K^+)
= \frac{0.01741 (\sqrt{s}-1.688)^{1.2078}}{(\sqrt{s}-1.725)^2+0.003777}
\hspace{0.5cm} {\rm mb},  \label{prmtpids4}
\end{equation}
\\
The parametrizations for the cross sections
$\sigma(\pi \Delta \rightarrow \Sigma K)$
in eqs. (\ref{prmtpids1}) to (\ref{prmtpids4}) should be understood
to be zero below the threshold ($\sqrt{s} \leq 1.688$ GeV).
These parametrizations are useful for codes which simulate
kaon production since no experimental data are available.

\subsection{Summary}\

In this paper, we have calculated the $\pi N \rightarrow Y K$
and $\pi \Delta \rightarrow Y K$ cross sections using a resonance
model on the hadronic
level.  All resonances with known experimental branching ratios
to  $\pi N$ and $Y K$ are included.

We constructed the interaction Lagrangians needed in the description of
the vertices.

For the reactions $\pi N \rightarrow \Lambda  K$ our theoretical calculations
are in reasonable good agreement with the experimental data both for
differential and total cross sections.
For the  $\pi N \rightarrow \Sigma  K$  reactions, we scaled the
coupling constant related to the $\Delta(1920)$ resonance, to take
into account also other  $\Delta$ resonances in this energy region.
To get a complete understanding of these reactions a better determination
of the branching ratios also of other $\Delta$ resonances is necessary.
For the  $\pi \Delta \rightarrow Y K$  reactions, we predict the total
cross sections and find that the differential cross sections
are almost isotopic.

For all different reaction channels with different charges of
the particapants  the total cross sections are
parameterized for the application in heavy ion collisions  \\

\noindent {\bf Acknowledgement:} The authors  express their
thanks to Prof. R. Vinh Mau, Prof. H. M\"uther and Prof. E. Oset for useful
discussions and are indebted to Prof. K. W. Schmid for providing
the code used for fitting the theoretically calculated cross
sections to the simple parametrisations given in this work.

\newpage
\section{Appendix}\
In this appendix, we list all the theoretical branching ratios of
the resonances used in this work.

\subsection{Branching ratios for $\pi+N \rightarrow \Lambda +K $ reactions}\

The branching ratios averaged over initial and summed over final
spin and isospin states of the baryon resonances and $K^*(892)$ are given:
(form factors have to be inserted at each vertex see $\S$ 3.1)
\begin{equation}
\Gamma(N(1650) \rightarrow N \pi) =
3 \frac{g^2_{\pi N N(1650)} }{4\pi} \frac{( \sqrt {m_N^2+
{\vec p}_N^{\hspace{1mm} 2} } + m_N)}{m_{N(1650)}}
|{\vec p}_N|,
\label{ratio511}
\end{equation}
with $ |{\vec p}_N|=\frac{\lambda^{\frac{1}{2}}
(m_{N(1650)}^2,m_N^2,m_{\pi}^2)}{2m_{N(1650)}}$

\begin{equation}
\Gamma(N(1710) \rightarrow N \pi) =
3 \frac{g^2_{\pi N N(1710)} }{4\pi} \frac{( \sqrt
{m_N^2+{\vec p}_N^{\hspace{1mm}2}} - m_N)}{m_{N(1710)}}
|{\vec p}_N|,
\label{ratio512}
\end{equation}
with $|{\vec p}_N|=\frac{\lambda^{\frac{1}{2}}
(m_{N(1710)}^2,m_N^2,m_{\pi}^2)}{2m_{N(1710)}}$

\begin{equation}
\Gamma(N(1720) \rightarrow N \pi) =
3 \frac{g^2_{\pi N N(1720)} }{12\pi m_{N(1720)} m_{\pi}^2  }
\frac{( \sqrt {m_N^2+{\vec p}_N^{\hspace{1mm}2}} + m_N)}{m_{N(1720)}}
|{\vec p}_N|,
\label{ratio513}
\end{equation}
with $|{\vec p}_N|=\frac{\lambda^{\frac{1}{2}}
(m_{N(1720)}^2,m_N^2,m_{\pi}^2)}{2m_{N(1720)}}$

\begin{equation}
\Gamma(N(1650) \rightarrow \Lambda K) =
 \frac{g^2_{\Lambda K N(1650)} }{4\pi}
 \frac{(\sqrt {m_\Lambda^2+{\vec p}_\Lambda^{\hspace{1mm}2}} +
 m_{\Lambda})}{m_{N(1650)}}
|{\vec p}_\Lambda|,
\label{ratio514}
\end{equation}
with $|{\vec p}_\Lambda|=\frac{\lambda^{\frac{1}{2}}
(m_{N(1650)}^2,m_\Lambda^2,m_K^2)}{2m_{N(1650)}}$

\begin{equation}
\Gamma(N(1710) \rightarrow {\Lambda} K) =
\frac{g^2_{{\Lambda} K N(1710)} }{4\pi}
\frac{(\sqrt {m_\Lambda^2+{\vec p}_\Lambda^{\hspace{1mm}2}}-
m_{\Lambda})}{m_{N(1710)}}
|{\vec p}_\Lambda|,
\label{ratio515}
\end{equation}
with $|{\vec p}_\Lambda|=\frac{\lambda^{\frac{1}{2}}
(m_{N(1710)}^2,m_\Lambda^2,m_K^2)}{2m_{N(1710)}}$

\begin{equation}
\Gamma(N(1720) \rightarrow \Lambda K) =
\frac{g^2_{K \Lambda N(1720)} }{12\pi m_{N(1720)} m_K^2  }
\frac{(\sqrt {m_\Lambda^2+{\vec p}_\Lambda^{\hspace{1mm}2}} +
m_{\Lambda} )}{m_{N(1720)}}
|{\vec p}_\Lambda|,
\label{ratio516}
\end{equation}
with $|{\vec p}_\Lambda|=\frac{\lambda^{\frac{1}{2}}
(m_{N(1720)}^2,m_\Lambda^2,m_K^2)}{2m_{N(1720)}}$

\begin{equation}
\Gamma(K^*(892) \rightarrow K \pi) =
3 \frac{f^2_{K^*(892) K \pi}}{4\pi}
\frac{2}{3 m^2_{K^*(892)}}|{\vec p}_K|^3 ,
\label{ratio517}
\end{equation}
with $|{\vec p}_K|=\frac{\lambda^{\frac{1}{2}}
(m_{K^*(892)}^2,m_K^2,m_\pi^2)}{2m_{K^*(892)}}$

%
\subsection{Branching ratios for $\pi+N \rightarrow \Sigma +K$  reactions}\

Besides eqs. (23)-(24) and eq. (28), the decay width  needed in determining
the effective coupling constants are given by:

\begin{equation}
\Gamma(\Delta (1920) \rightarrow N \pi) =
\frac{g^2_{\pi N \Delta(1920) }}{12\pi}
\frac{(\sqrt{m_N^2+{\vec p}_N^{\hspace{1mm}2}}+
m_N)}{m_{\Delta(1920)} m_{\pi}^2}
|{\vec p}_N|^3
\label{ratio521},
\end{equation}
with $|{\vec p}_N|=\frac{\lambda^{\frac{1}{2}}
(m_{\Delta(1920)}^2,m_N^2,m_{\pi}^2)}{2m_{\Delta(1920)}}$

\begin{equation}
\Gamma(N(1710) \rightarrow \Sigma K) =
3 \frac{g^2_{K \Sigma N(1710)}  }{4\pi}
\frac{(\sqrt{m_\Sigma^2+{\vec p}_\Sigma^{\hspace{1mm}2}} - m_\Sigma)}
{m_{N(1710)}}
|{\vec p}_\Sigma|,
\label{ratio522}
\end{equation}
with $|{\vec p}_\Sigma|=\frac{\lambda^{\frac{1}{2}}
(m_{N(1710)}^2,m_\Sigma^2,m_K^2)}{2m_{N(1710)}}$

\begin{equation}
\Gamma(N(1720) \rightarrow \Sigma K) =
3 \frac{g^2_{K \Sigma N(1720)}  }{12\pi}
\frac{(\sqrt{m_\Sigma^2+{\vec p}_\Sigma^{\hspace{1mm}2}} + m_\Sigma)}
{m_{N(1720)} m_K^2}
|{\vec p}_\Sigma|^3,
\label{ratio523}
\end{equation}
with $|{\vec p}_\Sigma|=\frac{\lambda^{\frac{1}{2}}
(m_{N(1720)}^2,m_\Sigma^2,m_K^2)}{2m_{N(1720)}}$

\begin{equation}
\Gamma(\Delta(1920) \rightarrow \Sigma K) =
\frac{g^2_{K \Sigma \Delta(1920)}  }{12\pi}
\frac{(\sqrt{m_\Sigma^2+{\vec p}_\Sigma^{\hspace{1mm}2}} + m_\Sigma)}
{m_{\Delta (1920)}m_K^2}
|{\vec p}_\Sigma|^3,
\label{ratio524}
\end{equation}
with $|{\vec p}_\Sigma|=\frac{\lambda^{\frac{1}{2}}
(m_{\Delta(1920)}^2,m_\Sigma^2,m_K^2)}{2m_{\Delta(1920)}}.$

\subsection{Branching ratios for $\pi+\Delta(1232) \rightarrow
\Lambda +K $ reactions}\

\begin{equation}
\Gamma(N(1650) \rightarrow \Delta \pi) =
2 \frac{g^2_{\pi \Delta N(1650)}}{6\pi}
\frac{m_{N(1650)} (E_\Delta - m_\Delta)}{m_\pi^2 m_\Delta^2}
|{\vec p}_\Delta|^3,
\label{ratio531}
\end{equation}
with $|{\vec p}_\Delta|=\frac{\lambda^{\frac{1}{2}}
(m_{N(1650)}^2,m_\Delta^2,m_\pi^2)}{2m_{N(1650)}}$
\begin{equation}
\Gamma(N(1710) \rightarrow \Delta \pi) =
2 \frac{g^2_{\pi \Delta N(1710)} }{6\pi}
\frac{m_{N(1710)} (E_\Delta + m_\Delta)}{m_\pi^2 m_\Delta^2}
|{\vec p}_\Delta|^3,
\label{ratio532}
\end{equation}
with $|{\vec p}_\Delta|=\frac{\lambda^{\frac{1}{2}}
(m_{N(1710)}^2,m_\Delta^2,m_\pi^2)}{2m_{N(1710)}}$
$$
\Gamma(N(1720) \rightarrow \Delta \pi) =
2 \frac{g^2_{\pi \Delta N(1720)}}{36\pi}
|{\vec p}_\Delta|\hspace{2em}
$$
\begin{equation}
\hspace{14em}
\cdot (\frac{m_\Delta}{m_{N(1720)}}) \left[\,
(\frac{E_\Delta}{m_\Delta}) - 1\,\right]
\left[\,2(\frac{E_\Delta}{m_\Delta})^2 -
2(\frac{E_\Delta}{m_\Delta}) + 5\,\right] ,
\label{ratio533}
\end{equation}
with $|{\vec p}_\Delta|=\frac{\lambda^{\frac{1}{2}}
(m_{N(1720)}^2,m_\Delta^2,m_\pi^2)}{2m_{N(1720)}}.$
\vfill
\eject

\newpage
\begin{table}
\caption{Calculated coupling constants}
\begin{center}
\begin{tabular}{cccccccc}
\hline
$B^*$(resonance) & $\Gamma^{full}(MeV)$ & $\Gamma_{N \pi}(\%) $
& $g_{\pi N B^*}^2$ & $\Gamma_{\Lambda K}(\%)$ & $ g_{K \Lambda  B^*}^2 $ \\
\hline \\
$N(1650)$       & 150 & 70.0 & 1.41    & 7.0     & 6.40 $\times 10^{-1}$ \\
$N(1710)$       & 100 & 15.0 & 2.57    & 15.0    & 4.74 $\times 10^{+1}$ \\
$N(1720)$       & 150 & 15.0 & 5.27 $\times 10^{-2}$ & 6.5 & 3.91 \\
\hline
                & $f_{K^*(892) K \pi}^2$ & & $g_{K^*(892) \Lambda N }^2$ & \\
\hline
                & 6.89 $\times 10^{-1}$ &    & 2.03 $\times 10^{-1}$ & &   \\
                &($ \Gamma$=50 MeV,&$\Gamma_{K \pi}=100\% $) &   & & \\

\\
\hline
\end{tabular}
\end{center}
\end{table}
\vspace{1.5cm}
\noindent
{\bf Table 1}
\newline
The calculated coupling constants and the experimental branching
ratios used for the calculations of the coupling constants.
\\
%
\newpage
\begin{table}
\caption{Calculated and fitted coupling constants}
\begin{center}
\begin{tabular}{cccccc}
\hline
$B^*$(resonance) &$\Gamma^{full}(MeV)$
&$\Gamma_{N \pi}(\%)$       &${g_{\pi N B^*}^2}$
&$\Gamma_{\Sigma K}(\%)$    &${g_{K\Sigma B^*}^2}$  \\
\hline \\
$N(1710)$     &100 &15.0 &(see table 1)&6.0 &$4.5\times10^{+1}$\\
$N(1720)$     &150 &15.0 &(see table 1)&3.5 &
3.15\\
$\Delta(1920)$ (set 1)
&--&--&(1.44)&--&(3.83)\\
$\Delta(1920)$ (set 2)
&200 &12.5 &$4.17\times10^{-1}$&2.0 &1.11\\
\\
\hline
&${f_{K^*(892) K \pi}^2}$&
&${g_{K^*(892) \Sigma N}^2}$& &\\
\hline
\\
&(see table 1)& &$2.03\times10^{-1}$(=$g_{K^*(892) \Lambda N}^2$)&&\\
\\
\hline
\end{tabular}
\end{center}
\end{table}
\vspace{1.5cm}
\noindent
{\bf Table 2}
\newline
The calculated or fitted coupling constants and the data used for the
calculations.
The values in brackets stand for the coupling constants
obtained by fitting to the total cross section for the $\pi^+ p
\rightarrow \Sigma^+ K^+$ reaction (set 1).
The value of $\frac{g_{K^*(892) \Sigma N}^2}{4 \pi}$ is the
fitted value to the $\pi^+ p \rightarrow \Sigma^+ K^+$ channel
when zero tensor coupling for $K^*(892) \Sigma N$ interaction ($\kappa =0$) is
applied and the  rescaled coupling constants for $\Delta(1920)$ are used.

\newpage
\begin{table}
\caption{Calculated coupling constants}
\begin{center}
\begin{tabular}{cccccccc}
\hline
$B^*$(resonance) & $\Gamma^{full}(MeV)$ & $\Gamma_{\Delta \pi}(\%) $
& $g_{\pi \Delta B^*}^2$ & $\Gamma_{\Lambda K}(\%)$ & $ g_{K \Lambda B^*}^2 $\\
\hline \\
$N(1650)$       & 150 &  5.0 & 6.56 $\times 10^{-1}$ & 7.0   & (see table 1) \\
$N(1710)$       & 100 & 17.5 & 1.85 $\times 10^{-2}$ & 15.0  & (see table 1) \\
$N(1720)$       & 150 & 10.0 & 1.12 $\times 10^{+1}$ & 6.5   & (see table 1) \\
\hline
\\
\hline
\end{tabular}
\end{center}
\end{table}
\vspace{1.5cm}
\noindent
{\bf Table 3}
\newline
The calculated coupling constants and the experimental branching
ratios used for the calculations of the coupling constants.
\\
%

\newpage
\noindent
{\bf Figure captions}
\vspace{1.5cm}

\noindent
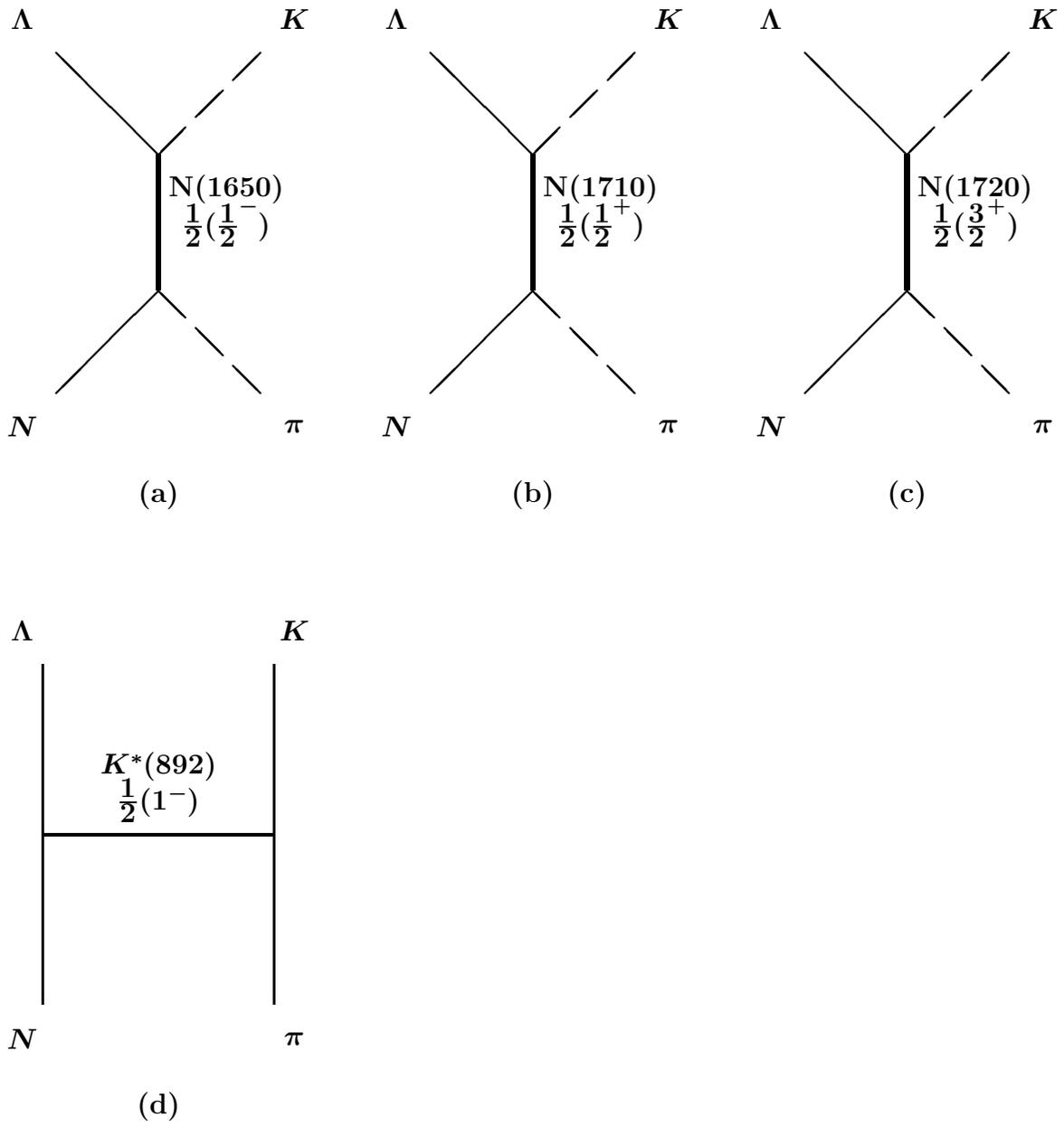
{\bf Fig. 1}
\newline
The processes contributing to the $\pi N \rightarrow \Lambda K$
reactions. The diagrams are corresponding to the different
intermediate resonance states:
$(a):\,\, N(1650)\, I(J^P)=\frac{1}{2}(\frac{1}{2}^-)$
$(b):\,\, N(1710)\, \frac{1}{2}(\frac{1}{2}^+)$,
$(c):\,\, N(1720)\, \frac{1}{2}(\frac{3}{2}^+)$ and
$(d):\,\,$t-channel $K^*(892)$-exchange, respectively.
\vspace{1cm}

\noindent
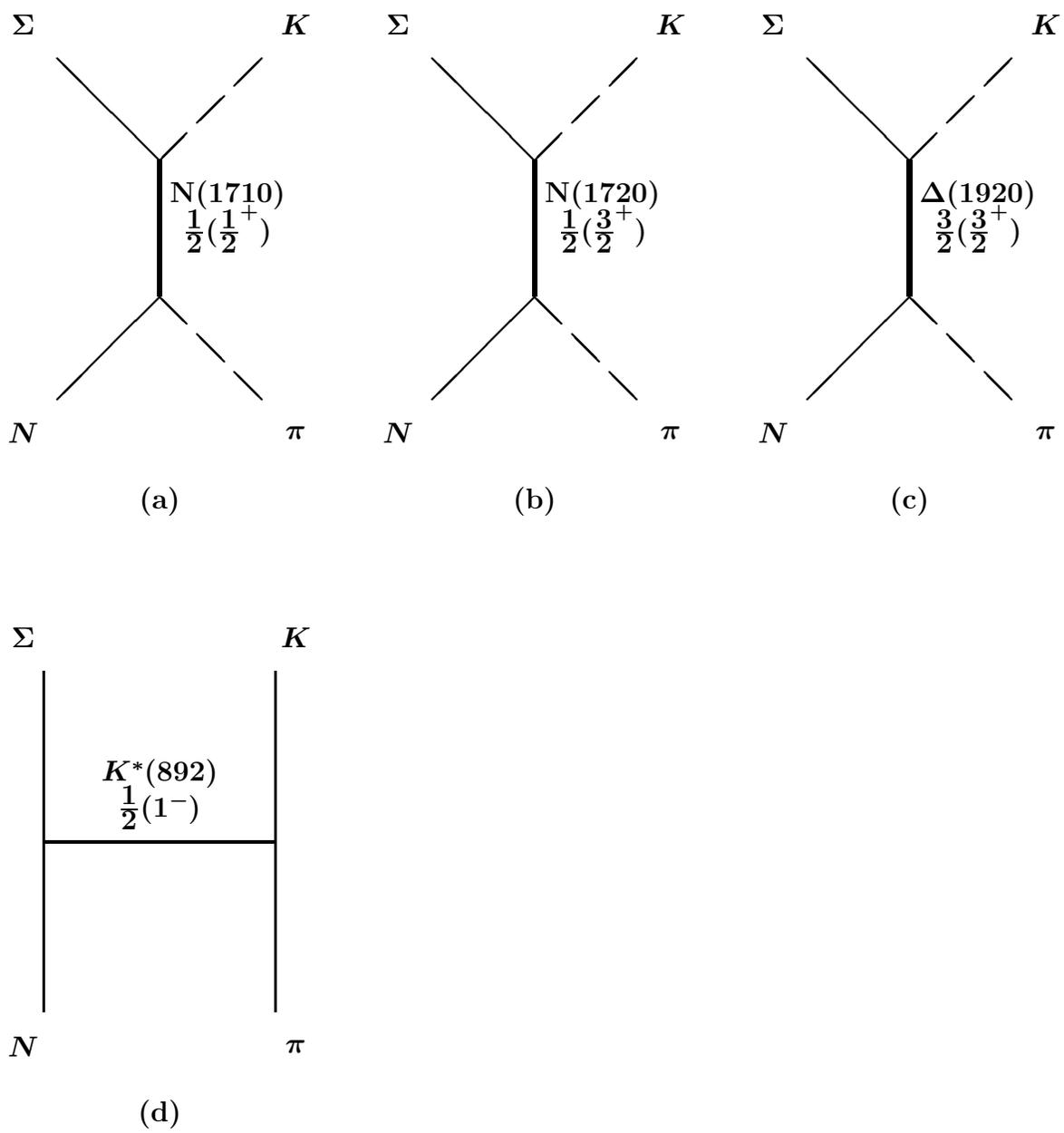
{\bf Fig. 2}
\newline
The processes contributing to the $\pi N \rightarrow \Sigma K$
reactions. The diagrams are corresponding to the different
intermediate resonance states:
$(a):\,\, N(1710)\,I(J^P) = \frac{1}{2}(\frac{1}{2}^+)$,
$(b):\,\, N(1720)\, \frac{1}{2}(\frac{3}{2}^+)$,
$(c):\,\, \Delta(1920)\, \frac{3}{2}(\frac{3}{2}^+)$ and
$(d):\,\,$t-channel $K^*(892)$ exchange, respectively.
\vspace{1cm}

\noindent
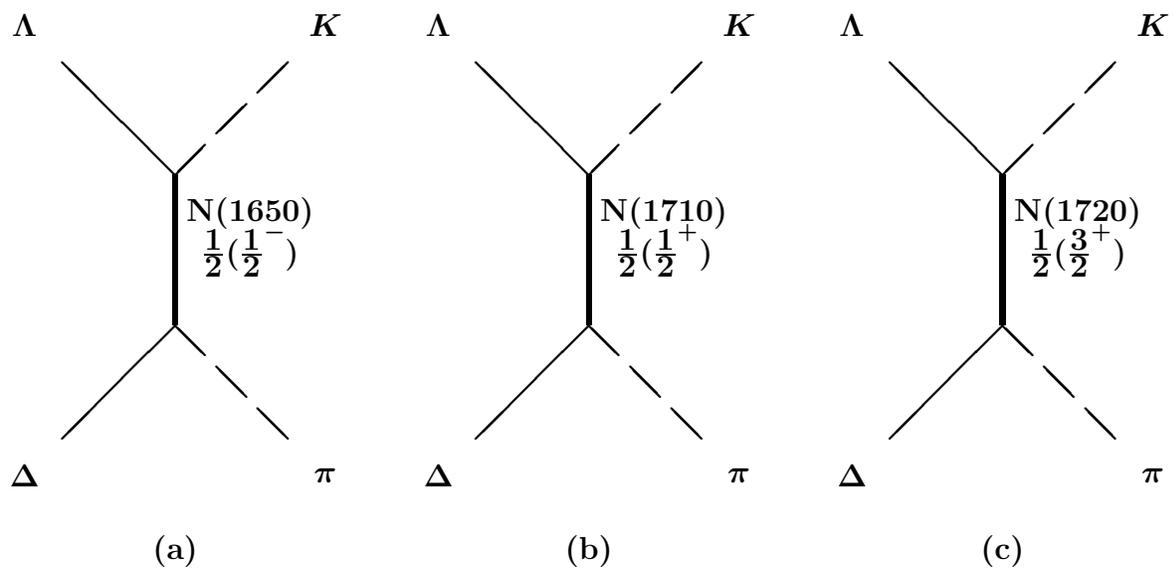
{\bf Fig. 3}
\newline
The processes contributing to the $\pi \Delta \rightarrow \Lambda K$
reactions. The diagrams are corresponding to the different
intermediate resonance states:
$(a):\,\, N(1650)\, I(J^P)=\frac{1}{2}(\frac{1}{2}^-)$
$(b):\,\, N(1710)\, \frac{1}{2}(\frac{1}{2}^+)$ and
$(c):\,\, N(1720)\, \frac{1}{2}(\frac{3}{2}^+)$,  respectively.
\vspace{1cm}

\noindent
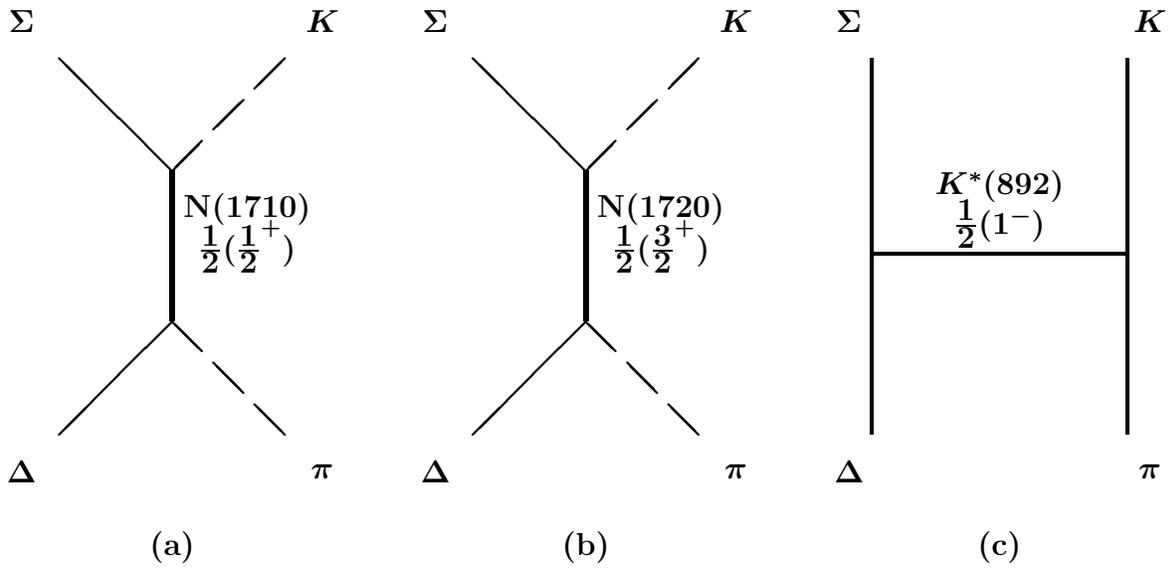
{\bf Fig. 4}
\newline
The processes contributing to the $\pi \Delta \rightarrow \Sigma K$
reactions. The diagrams are corresponding to the different
intermediate resonance states:
$(a):\,\, N(1710)\,I(J^P) = \frac{1}{2}(\frac{1}{2}^+)$,
$(b):\,\, N(1720)\, \frac{1}{2}(\frac{3}{2}^+)$ and
$(d):\,\,$t-channel $K^*(892)$ exchange, respectively.
\vspace{1cm}

\noindent
{\bf Fig. 5}
\newline
The total cross section
for the reaction $\pi^- p \rightarrow \Lambda K^0$
The experimental data with error bars are taken from \cite{bal}.
The theoretical calculation are shown with the solid lines for
the results without interference terms (fig. (a)) and
with interference terms (fig. (b)).
\vspace{1cm}

\noindent
{\bf Fig. 6}
\newline
The differential cross section for the reaction
$\pi^- p \rightarrow \Lambda K^0$ as a function
of $cos\theta$ in the c.m. frame at different bombarding energies.
(a),  (b) and (c) show the reactions at pion beam momenta
0.980 GeV/c (corresponding to $\sqrt s =1.66$ GeV), 1.13 GeV/c
($\sqrt s =1.742$ GeV)
and 1.455 GeV/c ($\sqrt s =1.908$ GeV), respectively.
The experimental data with error
bars for (a) and (b) are taken from \cite{kna75}.
The experimental data with error
bars for (c) are from \cite{sax80}.
The theoretical calculations (solid lines)
include interference terms.
\vspace{1cm}

\noindent
{\bf Fig. 7 }
\newline
(a). The calculated and experimental \cite{bal} total cross sections for the
$\pi^+ p \rightarrow \Sigma^+ K^+$
($\pi^- n \rightarrow \Sigma^- K^0$) reaction.
The solid line and the dashed line stand for the set 1
and set 2 parameters, respectively. Set 1 are the relevant coupling
constants for $\Delta(1920)$ and $g_{K^*(892) K \pi}$ adjusted to the
cross section $\pi^+ p \rightarrow \Sigma^+ K^+$
with the $\Delta(1920)$ as an s-channel intermediate state resonance and other
coupling constants are determined from the branching ratios.
Set 2 uses the same value of $g_{K^*(892) K \pi}$ as that of set
1 and all other coupling constants determined from the branching ratios.
\newline
(b). The calculated total cross sections for the
$\pi^- p \rightarrow \Sigma^- K^+$
($\pi^+ n \rightarrow \Sigma^+ K^0$) reaction.
\newline
(c).
The calculated total cross sections for the
$\pi^+ n \rightarrow \Sigma^0 K^+$ and
$\pi^0 n \rightarrow \Sigma^- K^+$
($\pi^0 p \rightarrow \Sigma^+ K^0$ and
$\pi^- p \rightarrow \Sigma^0 K^0$) reactions.
\vspace{1cm}

\noindent
{\bf Fig. 8}
\newline
(a). The calculated total cross sections for the
$\pi^- \Delta^{++} \rightarrow \Sigma^0 K^+$
($\pi^+ \Delta^- \rightarrow \Sigma^0 K^0$) reactions.
The solid line and the dashed lines stand for the results without
and with the inclusion the interference terms, respectively.
Note that the largest and the smallest results are displayed for
the four possibilities arising from the possible signs of the
coupling constants and thus the interference terms.
\newline
(b). The calculated total cross sections for the
$\pi^0 \Delta^0 \rightarrow \Sigma^- K^+$
($\pi^0 \Delta^{+} \rightarrow \Sigma^+ K^0$) reactions.
\newline
(c). The calculated total cross sections for the
$\pi^+ \Delta^{0} \rightarrow \Sigma^0 K^+$
($\pi^- \Delta^+ \rightarrow \Sigma^0 K^0$) reactions.

\newpage
{\Large \bf Fig. 1}\\
\bf
\boldmath
\setlength{\unitlength}{1cm}
\begin{picture}(11,20) \thicklines
\put(0,11){\makebox(1,1){$N$}}
\put(0,17){\makebox(1,1){$\Lambda$}}
\put(2,10){\makebox(1,1){(a)}}
\put(4,11){\makebox(1,1){$\pi$}}
\put(4,17){\makebox(1,1){$K$}}
\put(3,14.5){\makebox(1,1){N(1650)}}
\put(3,14){\makebox(1,1)
{$\frac{\displaystyle 1}{\displaystyle 2}
(\frac{\displaystyle 1}{\displaystyle 2}^-)$}}
\put(2.5,13.5){\line(-1,-1){1.5}}
\multiput(2.5,13.5)(0.55,-0.55){3}{\line(1,-1){0.4}}
\put(2.48,13.5){\line(0,1){2}}
\put(2.5,13.5){\line(0,1){2}}
\put(2.52,13.5){\line(0,1){2}}
\put(2.5,15.5){\line(-1,1){1.5}}
\multiput(2.5,15.5)(0.55,0.55){3}{\line(1,1){0.4}}
\put(5.5,11){\makebox(1,1){$N$}}
\put(5.5,17){\makebox(1,1){$\Lambda$}}
\put(7.5,10){\makebox(1,1){(b)}}
\put(9.5,11){\makebox(1,1){$\pi$}}
\put(9.5,17){\makebox(1,1){$K$}}
\put(8.5,14.5){\makebox(1,1){N(1710)}}
\put(8.5,14){\makebox(1,1)
{$\frac{\displaystyle 1}{\displaystyle 2}
(\frac{\displaystyle 1}{\displaystyle 2}^+)$}}
\put(8,13.5){\line(-1,-1){1.5}}
\multiput(8,13.5)(0.55,-0.55){3}{\line(1,-1){0.4}}
\put(7.98,13.5){\line(0,1){2}}
\put(8,13.5){\line(0,1){2}}
\put(8.02,13.5){\line(0,1){2}}
\put(8,15.5){\line(-1,1){1.5}}
\multiput(8,15.5)(0.55,0.55){3}{\line(1,1){0.4}}
\put(11,11){\makebox(1,1){$N$}}
\put(11,17){\makebox(1,1){$\Lambda$}}
\put(13,10){\makebox(1,1){(c)}}
\put(15,11){\makebox(1,1){$\pi$}}
\put(15,17){\makebox(1,1){$K$}}
\put(14,14.5){\makebox(1,1){N(1720)}}
\put(14,14){\makebox(1,1)
{$\frac{\displaystyle 1}{\displaystyle 2}
(\frac{\displaystyle 3}{\displaystyle 2}^+)$}}
\put(13.5,13.5){\line(-1,-1){1.5}}
\multiput(13.5,13.5)(0.55,-0.55){3}{\line(1,-1){0.4}}
\put(13.48,13.5){\line(0,1){2}}
\put(13.5,13.5){\line(0,1){2}}
\put(13.52,13.5){\line(0,1){2}}
\put(13.5,15.5){\line(-1,1){1.5}}
\multiput(13.5,15.5)(0.55,0.55){3}{\line(1,1){0.4}}
\put(0,2){\makebox(1,1){$N$}}
\put(0,8){\makebox(1,1){$\Lambda$}}
\put(2,1){\makebox(1,1){(d)}}
\put(4,2){\makebox(1,1){$\pi$}}
\put(4,8){\makebox(1,1){$K$}}
\put(2,6){\makebox(1,1){$K^*$(892)}}
\put(2,5.5){\makebox(1,1)
{$\frac{\displaystyle 1}{\displaystyle 2}({\displaystyle 1}^-)$}}
\put(0.8,3){\line(0,1){5}}
\put(4.2,3){\line(0,1){5}}
\put(0.8,5.5){\line(1,0){3.4}}
\end{picture}
\newpage
{\Large \bf Fig. 2}\\
\bf
\boldmath
\setlength{\unitlength}{1cm}
\begin{picture}(11,20) \thicklines
\put(0,11){\makebox(1,1){$N$}}
\put(0,17){\makebox(1,1){$\Sigma$}}
\put(2,10){\makebox(1,1){(a)}}
\put(4,11){\makebox(1,1){$\pi$}}
\put(4,17){\makebox(1,1){$K$}}
\put(3,14.5){\makebox(1,1){N(1710)}}
\put(3,14){\makebox(1,1)
{$\frac{\displaystyle 1}{\displaystyle 2}
(\frac{\displaystyle 1}{\displaystyle 2}^+)$}}
\put(2.5,13.5){\line(-1,-1){1.5}}
\multiput(2.5,13.5)(0.55,-0.55){3}{\line(1,-1){0.4}}
\put(2.48,13.5){\line(0,1){2}}
\put(2.5,13.5){\line(0,1){2}}
\put(2.52,13.5){\line(0,1){2}}
\put(2.5,15.5){\line(-1,1){1.5}}
\multiput(2.5,15.5)(0.55,0.55){3}{\line(1,1){0.4}}
\put(5.5,11){\makebox(1,1){$N$}}
\put(5.5,17){\makebox(1,1){$\Sigma$}}
\put(7.5,10){\makebox(1,1){(b)}}
\put(9.5,11){\makebox(1,1){$\pi$}}
\put(9.5,17){\makebox(1,1){$K$}}
\put(8.5,14.5){\makebox(1,1){N(1720)}}
\put(8.5,14){\makebox(1,1)
{$\frac{\displaystyle 1}{\displaystyle 2}
(\frac{\displaystyle 3}{\displaystyle 2}^+)$}}
\put(8,13.5){\line(-1,-1){1.5}}
\multiput(8,13.5)(0.55,-0.55){3}{\line(1,-1){0.4}}
\put(7.98,13.5){\line(0,1){2}}
\put(8,13.5){\line(0,1){2}}
\put(8.02,13.5){\line(0,1){2}}
\put(8,15.5){\line(-1,1){1.5}}
\multiput(8,15.5)(0.55,0.55){3}{\line(1,1){0.4}}
\put(11,11){\makebox(1,1){$N$}}
\put(11,17){\makebox(1,1){$\Sigma$}}
\put(13,10){\makebox(1,1){(c)}}
\put(15,11){\makebox(1,1){$\pi$}}
\put(15,17){\makebox(1,1){$K$}}
\put(14,14.5){\makebox(1,1){$\Delta$(1920)}}
\put(14,14){\makebox(1,1)
{$\frac{\displaystyle 3}{\displaystyle 2}
(\frac{\displaystyle 3}{\displaystyle 2}^+)$}}
\put(13.5,13.5){\line(-1,-1){1.5}}
\multiput(13.5,13.5)(0.55,-0.55){3}{\line(1,-1){0.4}}
\put(13.48,13.5){\line(0,1){2}}
\put(13.5,13.5){\line(0,1){2}}
\put(13.52,13.5){\line(0,1){2}}
\put(13.5,15.5){\line(-1,1){1.5}}
\multiput(13.5,15.5)(0.55,0.55){3}{\line(1,1){0.4}}
\put(0,2){\makebox(1,1){$N$}}
\put(0,8){\makebox(1,1){$\Sigma$}}
\put(2,1){\makebox(1,1){(d)}}
\put(4,2){\makebox(1,1){$\pi$}}
\put(4,8){\makebox(1,1){$K$}}
\put(2,6){\makebox(1,1){$K^*$(892)}}
\put(2,5.5){\makebox(1,1)
{$\frac{\displaystyle 1}{\displaystyle 2}({\displaystyle 1}^-)$}}
\put(0.8,3){\line(0,1){5}}
\put(4.2,3){\line(0,1){5}}
\put(0.8,5.5){\line(1,0){3.4}}
\end{picture}
\newpage
{\Large \bf Fig. 3}\\
\bf
\boldmath
\setlength{\unitlength}{1cm}
\begin{picture}(11,20) \thicklines
\put(0,11){\makebox(1,1){$\Delta$}}
\put(0,17){\makebox(1,1){$\Lambda$}}
\put(2,10){\makebox(1,1){(a)}}
\put(4,11){\makebox(1,1){$\pi$}}
\put(4,17){\makebox(1,1){$K$}}
\put(3,14.5){\makebox(1,1){N(1650)}}
\put(3,14){\makebox(1,1)
{$\frac{\displaystyle 1}{\displaystyle 2}
(\frac{\displaystyle 1}{\displaystyle 2}^-)$}}
\put(2.5,13.5){\line(-1,-1){1.5}}
\multiput(2.5,13.5)(0.55,-0.55){3}{\line(1,-1){0.4}}
\put(2.48,13.5){\line(0,1){2}}
\put(2.5,13.5){\line(0,1){2}}
\put(2.52,13.5){\line(0,1){2}}
\put(2.5,15.5){\line(-1,1){1.5}}
\multiput(2.5,15.5)(0.55,0.55){3}{\line(1,1){0.4}}
\put(5.5,11){\makebox(1,1){$\Delta$}}
\put(5.5,17){\makebox(1,1){$\Lambda$}}
\put(7.5,10){\makebox(1,1){(b)}}
\put(9.5,11){\makebox(1,1){$\pi$}}
\put(9.5,17){\makebox(1,1){$K$}}
\put(8.5,14.5){\makebox(1,1){N(1710)}}
\put(8.5,14){\makebox(1,1)
{$\frac{\displaystyle 1}{\displaystyle 2}
(\frac{\displaystyle 1}{\displaystyle 2}^+)$}}
\put(8,13.5){\line(-1,-1){1.5}}
\multiput(8,13.5)(0.55,-0.55){3}{\line(1,-1){0.4}}
\put(7.98,13.5){\line(0,1){2}}
\put(8,13.5){\line(0,1){2}}
\put(8.02,13.5){\line(0,1){2}}
\put(8,15.5){\line(-1,1){1.5}}
\multiput(8,15.5)(0.55,0.55){3}{\line(1,1){0.4}}
\put(11,11){\makebox(1,1){$\Delta$}}
\put(11,17){\makebox(1,1){$\Lambda$}}
\put(13,10){\makebox(1,1){(c)}}
\put(15,11){\makebox(1,1){$\pi$}}
\put(15,17){\makebox(1,1){$K$}}
\put(14,14.5){\makebox(1,1){N(1720)}}
\put(14,14){\makebox(1,1)
{$\frac{\displaystyle 1}{\displaystyle 2}
(\frac{\displaystyle 3}{\displaystyle 2}^+)$}}
\put(13.5,13.5){\line(-1,-1){1.5}}
\multiput(13.5,13.5)(0.55,-0.55){3}{\line(1,-1){0.4}}
\put(13.48,13.5){\line(0,1){2}}
\put(13.5,13.5){\line(0,1){2}}
\put(13.52,13.5){\line(0,1){2}}
\put(13.5,15.5){\line(-1,1){1.5}}
\multiput(13.5,15.5)(0.55,0.55){3}{\line(1,1){0.4}}
\end{picture}
\newpage
{\Large \bf Fig. 4}\\
\bf
\boldmath
\setlength{\unitlength}{1cm}
\begin{picture}(11,20) \thicklines
\put(0,11){\makebox(1,1){$\Delta$}}
\put(0,17){\makebox(1,1){$\Sigma$}}
\put(2,10){\makebox(1,1){(a)}}
\put(4,11){\makebox(1,1){$\pi$}}
\put(4,17){\makebox(1,1){$K$}}
\put(3,14.5){\makebox(1,1){N(1710)}}
\put(3,14){\makebox(1,1)
{$\frac{\displaystyle 1}{\displaystyle 2}
(\frac{\displaystyle 1}{\displaystyle 2}^+)$}}
\put(2.5,13.5){\line(-1,-1){1.5}}
\multiput(2.5,13.5)(0.55,-0.55){3}{\line(1,-1){0.4}}
\put(2.48,13.5){\line(0,1){2}}
\put(2.5,13.5){\line(0,1){2}}
\put(2.52,13.5){\line(0,1){2}}
\put(2.5,15.5){\line(-1,1){1.5}}
\multiput(2.5,15.5)(0.55,0.55){3}{\line(1,1){0.4}}
\put(5.5,11){\makebox(1,1){$\Delta$}}
\put(5.5,17){\makebox(1,1){$\Sigma$}}
\put(7.5,10){\makebox(1,1){(b)}}
\put(9.5,11){\makebox(1,1){$\pi$}}
\put(9.5,17){\makebox(1,1){$K$}}
\put(8.5,14.5){\makebox(1,1){N(1720)}}
\put(8.5,14){\makebox(1,1)
{$\frac{\displaystyle 1}{\displaystyle 2}
(\frac{\displaystyle 3}{\displaystyle 2}^+)$}}
\put(8,13.5){\line(-1,-1){1.5}}
\multiput(8,13.5)(0.55,-0.55){3}{\line(1,-1){0.4}}
\put(7.98,13.5){\line(0,1){2}}
\put(8,13.5){\line(0,1){2}}
\put(8.02,13.5){\line(0,1){2}}
\put(8,15.5){\line(-1,1){1.5}}
\multiput(8,15.5)(0.55,0.55){3}{\line(1,1){0.4}}
\put(11,11){\makebox(1,1){$\Delta$}}
\put(11,17){\makebox(1,1){$\Sigma$}}
\put(13,10){\makebox(1,1){(c)}}
\put(15,11){\makebox(1,1){$\pi$}}
\put(15,17){\makebox(1,1){$K$}}
\put(13,14.8){\makebox(1,1){$K^*$(892)}}
\put(13,14.3){\makebox(1,1)
{$\frac{\displaystyle 1}{\displaystyle 2}({\displaystyle 1}^-)$}}
\put(11.8,12){\line(0,1){5}}
\put(15.2,12){\line(0,1){5}}
\put(11.8,14.4){\line(1,0){3.4}}
\end{picture}

\begin{thebibliography}{99}
%
\bibitem{aic}J. Aichelin and C. M. Ko, Phys. Rev. Lett. 55  2661 (1985) .
%
\bibitem{lan}A. Lang, W. Cassing, U. Mosel and K. Weber, Nucl. Phys. A 541
507 (1992); S. W. Huang, A. Faessler, G. Q. Li, R. K. Puri,
E. Lehmann, D. T. Khoa and M. A. Matin, Phys. Lett. B 298 41 (1993); G.
Hartnack, J. J\"anicke and J. Aichelin, preprint, Universite de
Nantes, Rapport Interne LPN-93-11 (1993).
%
\bibitem{cas1}W. Cassing, V. Metag, U. Mosel and K. Niita,
Phys. Rep. 188 363 (1990).
%
\bibitem{li}G. Q. Li, A. Faessler and S. W. Huang, Progresses in
particles and nuclear physics 31 159 (1993).
%
%
%
\bibitem{bert}G. F. Bertsch, H. Kruse, S. Das Gupta, Phys. Rev.
{\bf C29} 673 (1984); H. Kruse, B. V. Jacak, H. St\"ocker, Phys. Rev. Lett.
{\bf 54} 289 (1985); G. F. Bertsch, S. Das Gupta, Phys. Rep.
{\bf 160} 189 (1988).
%
\bibitem{qmd}J. Aichelin, Phys. Rep. 202 235 (1991).
%
\bibitem{ran}J. Randrup and C. M. Ko, Nucl. Phys. A 343 519 (1980);
Nucl. Phys. A 411 537 (1983).
%
\bibitem{cug}J. Cugnon and R. M. Lombard, Nucl. Phys. A 422
635 (1984).
%
\bibitem{fer}E. Ferrari, Phys. Rev. 120 988 (1960);
T. Yao, Phys. Rev. 125 1048 (1962).
%
\bibitem{cas2} W. Cassing, G. Batko, U. Mosel,K. Niita, O.Schult
and Gy. Wolf, Phys. Lett. B 238 25 (1990).
%
\bibitem{wol}Gy. Wolf, G. Batko, W. Cassing, U. Mosel, K. Niita and
M. Sch\"afer, Nucl. Phys. A517 615 (1990).
%
\bibitem{hub}S. Huber and J. Aichelin, Nucl. Phys. A573 587 (1994).
%
\bibitem{tsu}K. Tsushima, S. W. Huang, A. Faessler,
Phys. Lett. B 337 245 (1994); K. Tsushima, S. W. Huang, A. Faessler,
J. Phys. G. in press.
%
\bibitem{par}Particle Data Group, Phys. Rev. D 45 (1992).
%
\bibitem{paro}Particle Data Group, Phys. Lett. B 239 1 (1990).
%
\bibitem{tak}See, for example, Y. Takahashi, {\it An
Introduction to Field Quantization}, Pergamon Press, 1969; D.
Luri\'e, {\it Particles and Fields}, John Wiley \& Sons, 1968.
%
\bibitem{ben}M. Benmerrouche, R. M. Davidson and Nimai C. Mukhopadhyay,
Phys. Rev. C 39 2339 (1989).
%
\bibitem{gob}C. Gobbi, F. Iachello and D. Kusnezov,
Yale University preprint (1994) PACS numbers: 13.25.+m,
11.30.Na, 12.40.Aa.
%
\bibitem{itz}C. Itzykson and J. B. Zuber, {\it Quantum Field Theory,}
McGraw-Hill, New York (1980).
%
\bibitem{bal}A. Baldini, V. Flamino, W. G. Moorhead and D. R. O.
Morrison, {\it Landolt-B\"ornstein, Numerical Data and
Functional Relationships in Science and Technology,} Vol. 12,
ed. by H. Schopper, (Springer-Verlag, 1988).
%
\bibitem{kna75} T. M. Knasel, et al. Phys. Rev. D 11 1 (1975)
%
\bibitem{sax80} D. H. Saxon et al. Nucl. Phys. B162 522 (1980).
%
\bibitem{bro}G. E. Brown, C. M. Ko, Z. G. Wu and L. H. Xia,
Phys. Rev. C 43 1881 (1991); H. W. Barz and H. Iwe, Nucl Phys. A
453 728 (1986); V. Pantuev, Phys. Lett. B 281 20 (1992).
%
%
%
%
%
%
%
\end{thebibliography}
\end{document}